\documentclass[prl, preprint, superscriptaddress, showkeys]{revtex4-2}

\usepackage{xr}
\usepackage{hyperref}
\hypersetup{colorlinks=true, citecolor=blue, urlcolor=blue, linkcolor=blue}

\usepackage{graphicx}
\usepackage{soul}
\usepackage{xcolor} 

\newcommand{\eb}[2]{#2}

\usepackage[hmargin=2cm,vmargin=2.5cm]{geometry}
\usepackage{filecontents}

\newcommand{\ens}{Laboratoire de Physique de l'Ecole normale sup\'erieure, ENS, Universit\'e PSL, CNRS, Sorbonne Universit\'e, Universit\'e de Paris, 24 rue Lhomond, 75005 Paris, France}
\newcommand{\trt}{Thales Research \& Technology, 91767 Palaiseau, France}
\newcommand{\fst}{Laboratoire de Physique de la Matière Condensée, Faculté des Sciences de Tunis, Université Tunis El Manar, Campus Universitaire 1060 Tunis, Tunisia.}
\newcommand{\fsb}{Laboratoire de Physique des Matériaux : Structure et Propriétés, Faculté des Sciences de Bizerte, Université de Carthage, 7021 Jarzouna, Tunisia.}
\newcommand{\iuf}{Institut universitaire de France}

\begin{document}

\title{
The optical absorption in indirect semiconductor to semimetal \texorpdfstring{PtSe\textsubscript{2}}{PtSe2} arises from direct transitions
}

\author{Marin Tharrault}
\affiliation{\ens}
\author{Sabrine Ayari}
\affiliation{\ens}
\author{Mehdi Arfaoui}
\affiliation{\fst}
\author{Eva Desgué}
\affiliation{\trt}
\author{Romaric Le Goff}
\affiliation{\ens}
\author{Pascal Morfin}
\affiliation{\ens}
\author{José Palomo}
\affiliation{\ens}
\author{Michael Rosticher}
\affiliation{\ens}
\author{Sihem Jaziri}
\affiliation{\fst}
\affiliation{\fsb}
\author{Bernard Plaçais}
\affiliation{\ens}
\author{Pierre Legagneux}
\affiliation{\trt}
\author{Francesca Carosella}
\affiliation{\ens}
\author{Christophe Voisin}
\affiliation{\ens}
\author{Robson Ferreira}
\affiliation{\ens}
\author{Emmanuel Baudin}
\affiliation{\ens}
\affiliation{\iuf}

\begin{abstract}
$\rm{PtSe_2}$ is a van der Waals material \eb{transiting}{transitioning} from an indirect \eb{}{bandgap} semiconductor to a semimetal with increasing thickness. Its absorption threshold has been conjectured to originate from interband indirect transitions. By quantitative comparison between \eb{wideband}{broadband} \eb{}{($0.8 - 3.0\,\rm{eV}$)} optical absorption \eb{($0.8 - 3.0\,\rm{eV}$)}{} of high-quality exfoliated crystals and \eb{}{DFT} \textit{ab initio} simulations, we prove instead that the optical absorption arises only from direct transitions. This understanding allows us to shed\eb{ new}{} light on the semiconductor-to-semimetal transition and to explore the effect of stacking and excitons on \eb{}{the} optical absorption.
\end{abstract}


\maketitle


Two-dimensional materials are promising candidates for the future of high-speed optoelectronics, owing to their large electronic mobilities, strong light-matter coupling, and intrinsically low carrier densities \cite{schmidtElectronic2015, Mak2016OptoTMD}. 
The near-infrared telecommunication band ($0.8 - 1.0\,\rm{eV}$) is particularly relevant to optoelectronic applications in information technology. However, only a few 2D materials can operate in this range, specifically the family of noble transition metal dichalcogenides \eb{}{(NTMDs)} \cite{Wang2022NTMDReviewPhotonics, LeGoff2021PdSe2}. 
Indeed, unlike other \eb{TMDs}{2D materials}, NTMDs exhibit a widely tunable bandgap with layer number, giving the possibility to target a desired wavelength range.
Among them, Platinum Diselenide ($\rm{PtSe_2}$) has raised significant interest due to its air stability, and to the availability of scalable methods to produce high-quality grown films \cite{wang2015monolayer, Wang2021ReviewPtSe2, Bonell2022MBE}.
$\rm{PtSe_2}$ features an indirect bandgap above $1\,\rm{eV}$ in its monolayer form and undergoes a transition to a semimetal as \eb{its thickness}{the number of stacked layers} increases \cite{Zhao2017SC_SM, Ciarrocchi2018SmScAndrasKis, Villaos2019DFT}. 
Yet, the light absorption mechanism on which photodetection processes rely is largely misunderstood.

Such issues are not uncommon when dealing with newly discovered materials. In the 1990s, there was \eb{}{an} intense debate regarding the nature of the optical bandgap of carbon nanotubes, which was eventually resolved by the demonstration of its excitonic origin \cite{Heinz2005CNT}.
In $\rm{PtSe_2}$, the \eb{low-energy}{near-infrared  (NIR)} absorption tail has been presumed to originate from indirect interband transitions \cite{Zhao2017SC_SM, Xie2019EllipsCVD, Chen2019PumpProbe, Shi2019CvdAu, Zhao2019NonLinAbs, Wang2019NLO, Gulo2020TempOpticalRaman, Fu2020PumpProbe, Ermolaev2021EllipsOscillator, Chung2021Hydrogen, Zheng2022PumpProbe, Su2023THz}  -- as it is the case for bulk indirect \eb{}{bandgap} semiconductors (e.g., Silicon) -- and some \eb{research}{studies} have hypothesized the involvement of bound\eb{s}{} excitons in its optical absorption peaks \cite{He2020absorptionExciton, He2022absorptionExciton, Bae2021PumpProbeExfo}.

In this letter, we demonstrate instead that the optical absorption of $\rm{PtSe_2}$ in the near-infrared to visible (NIR-vis) range is driven by single-electron direct optical transitions, irrespective of the material semimetallic or semiconducting nature.
Our conclusion is supported by a body of consistent experimental and theoretical evidence.
The mechanism of optical absorption is elucidated through quantitative agreement between micro-absorption spectroscopy of intrinsic exfoliated layer-defined $\rm{PtSe_2}$ flakes \eb{with}{and} DFT-based (density functional theory) optical absorption calculation, over the NIR-vis spectral range ($0.8 - 3.0\,\rm{eV}$). Alternative explanations are ruled out by studying the dependence of optical absorption with temperature and material quality.
Moreover, understanding NIR-vis optical absorption enables the determination of the number of layers for which $\rm{PtSe_2}$ transits from a semiconductor to a semimetal -- crucial for photodetection applications. 
It also allows us to investigate the impact of layer stacking and bound excitons on optical absorption.


\begin{figure*}[h]
\includegraphics[width=6.75in]{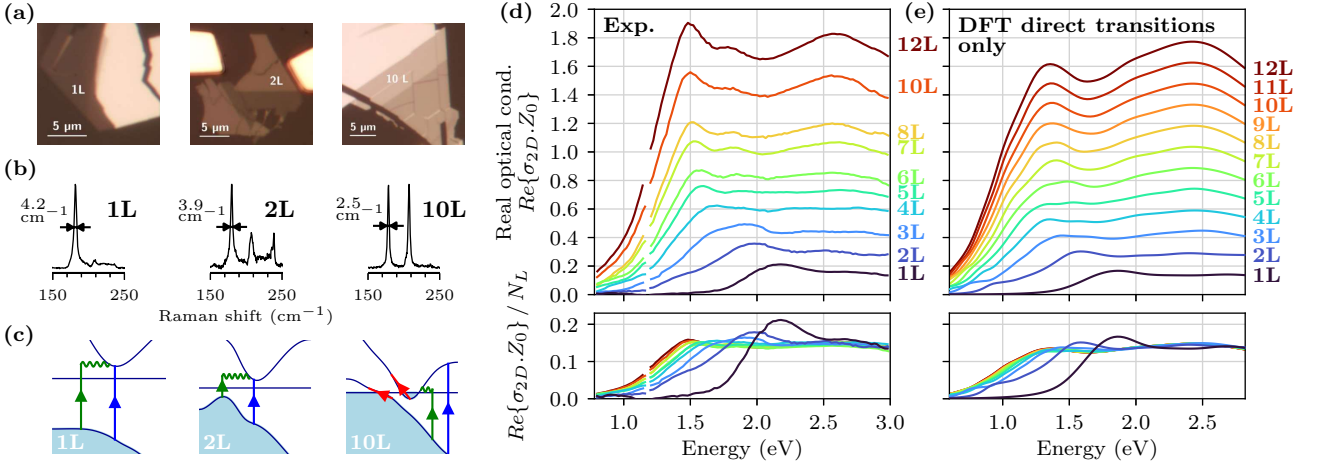}
\caption{Optical absorption of $\rm{PtSe_2}$ exfoliated flakes. (a) \eb{Optical images}{Microscope images on fused silica under white lamp illumination in reflection}, (b) Raman spectra \eb{}{(with $E_g$ mode linewidth)} and (c) \eb{}{schematic of} photon absorption mechanisms under consideration for 1L, 2L and 10L flakes \eb{}{(the horizontal line is the Fermi level)} -- depicting intraband (red), indirect interband (green) and direct interband (blue) transitions. (d) Experimental and (e) theoretical real 2D optical conductivities $Re\{\sigma_{2D}.Z_0\}$ (top plots), and their values normalized by the layers count $N_L$ (bottom plots). $Z_0 = 377\,\rm{\Omega}$ is the impedance of free space.}
\label{fig:conductivitiesSchemes}
\end{figure*}
$\rm{PtSe_2}$ samples are peeled from crystals grown by chemical vapor transport (HQ graphene) using Au-assisted mechanical exfoliation \cite{Desdai2016AuExfo, Heyl2020AuExfoAnneal}, and transferred onto fused silica substrates (Fig. \ref{fig:conductivitiesSchemes}a, \eb{}{see also }SM-\ref{SM-sec:AuAssistExfo}).
The monolayer to multilayers $\rm{PtSe_2}$ flakes feature the characteristic Raman signature\eb{}{s} of the \eb{1T}{octahedral} phase \cite{OBrien2016Raman} (Fig. \ref{fig:conductivitiesSchemes}b).
The $E_g$ Raman peaks have \eb{record }{}narrow linewidths, testifying of the high crystallinity of the exfoliated material.
These samples \eb{are subjected to}{have been the subject of} a detailed Raman study in reference \cite{articlePtSe2Raman}.


We use a high-accuracy \eb{V}{N}IR-vis ($0.8 - 3.0\,\rm{eV}$) micro-absorption setup at room temperature to study the exfoliated $\rm{PtSe_2}$ flakes\eb{ (SM-3)}{}. The acquisition of both reflectance and transmittance enables the determination of the complex two-dimensional optical conductivity \eb{(SM-3c)}{(SM-\ref{SM-sec:microAbsorption})}. For thin films, its real component (Fig. \ref{fig:conductivitiesSchemes}d) is simply the optical absorption in vacuum \cite{LiHeinz2018OptConduct}.
\eb{}{The optical spectra feature a plateau on the blue side ($2.5 - 3.0 \,\rm{eV}$, Fig. \ref{fig:conductivitiesSchemes}d), and show a slowly decaying tail on the infrared side.}
\eb{We identify}{This high-energy plateau allows identifying} the number of layers by using the linear scaling of the real 2D optical conductivity with layer count \eb{in the blue end of the electromagnetic spectrum ($2.5 - 3.0 \,\rm{eV}$, Fig. 1d)}{(Fig. \ref{fig:conductivitiesSchemes}d, bottom plot)}.
This \eb{attribution}{method} is consistent with atomic force microscopy thickness measurements (in combination with Raman spectroscopy, SM-\ref{SM-sec:AFM}).
\eb{Optical spectra feature a plateau on the blue side, and show a slowly decaying tail on the infrared side.}{}Two dominant peaks can be distinguished (at $1.5\,\rm{eV}$ and $2.6\,\rm{eV}$ for the thick samples), which redshift with increasing thickness. This effect is common in 2D semiconductors \cite{Zhao2013TMDExcitonShift}, but is particularly pronounced in $\rm{PtSe_2}$, due to its strong interlayer interaction.
The tail decay extends over an \eb{$0.5\,\rm{eV}$ }{}energy range \eb{-- much}{20 times} larger than \eb{}{the} thermal energy \eb{$\sim 26\,\rm{meV}$ -- which}{and so} cannot be attributed to thermal broadening of the main absorption peak. Interestingly, as the material transits from an indirect semiconductor to a semimetal with increasing thickness -- within the 3L - 10L range \cite{Ciarrocchi2018SmScAndrasKis, Yim2018TransportDC, Ansari2019ScSmGrown, Das2021ContactScSmPtSe2, Zhang2021STS, Li2021STM} -- no semimetallic signature emerges in the infrared \eb{end}{part} of the absorption spectrum.


Several mechanisms can be responsible for the NIR-vis optical absorption: intraband transitions, indirect interband transitions, and direct interband transitions (Fig. \ref{fig:conductivitiesSchemes}c) \cite{YuCardona2010ChapOptPtiesI}. 
\eb{}{Semimetallic $\rm{PtSe_2}$ shows free-electron signatures in the $\rm{THz}$ range \cite{Ji2023TDSThales, Su2023THz, HemmatTHz} but the absorption driven by intraband transitions can be neglected at optical frequencies.}

Owing to the indirect nature of the bandgap, and to the slow decay of the \eb{low-energy}{NIR} absorption tail, previous works studying optical absorption conjectured that indirect optical transitions are responsible for $\rm{PtSe_2}$ infrared absorption \cite{Zhao2017SC_SM, Xie2019EllipsCVD, Chen2019PumpProbe, Shi2019CvdAu, Zhao2019NonLinAbs, Wang2019NLO, Gulo2020TempOpticalRaman, Fu2020PumpProbe, Ermolaev2021EllipsOscillator, Chung2021Hydrogen, Zheng2022PumpProbe, Su2023THz}, and extracted indirect bandgaps using the Tauc model for indirect bandgap semiconductors \cite{Viezbicke2015TaucPlotZnO}.\eb{In the semimetal $\rm{PtSe_2}$, free-electron signatures are in the $\rm{THz}$ range [citations], and at optical frequencies, the absorption driven by intraband transitions can be neglected compared to the one due to direct transitions.}{}
\eb{Likewise, in}{However, it is} well-\eb{d}{kn}own \eb{}{that in} indirect semiconductors \eb{}{as} silicon and germanium, \eb{}{the high-energy} optical absorption\eb{ at high energy}{} is dominated by direct transitions, and orders of magnitude smaller indirect transitions are only revealed below the direct optical bandgap \cite{Dash1955SiGeOpticalAbsorption}.
We therefore compare the experimental spectra with the optical absorption deduced from DFT simulation, only considering direct optical transitions (Fig. \ref{fig:conductivitiesSchemes}e). DFT computation relies on the GGA-IPA method\eb{ (SM-4a)}{} for an AA-stacked \eb{$\rm{1T-PtSe_2}$}{octahedral $\rm{PtSe_2}$} structure\eb{}{~(SM-\ref{SM-sec:theory})}.
\eb{There is}{We obtain a} remarkable quantitative agreement between the \eb{}{simulated and measured} absorption magnitudes\eb{ of the simulation and the experiment}{.}\eb{, and the profile lineshapes}{ Additionally, the absorption profiles} feature similar absorption peaks and low-energy absorption tails. This striking correspondence \eb{supports}{strongly indicates} that the direct transition mechanism dominates the optical response. \eb{}{This interpretation is more thoroughly developed below.}


\eb{We now turn our discussion to}{Let us consider} the origin of \eb{}{the} optical absorption specific features, starting with the main absorption peaks. Several works \cite{He2022absorptionExciton, Bae2021PumpProbeExfo} attributed these peaks to bound excitons.
By analyzing the simulated optical absorption in detail, we see that these peaks originate instead from band nested direct transitions. \eb{We}{To demonstrate this point, let us} focus on the simplest cases of monolayer and bilayer $\rm{PtSe_2}$.

\begin{figure}[h]
\includegraphics[width=3.375in]{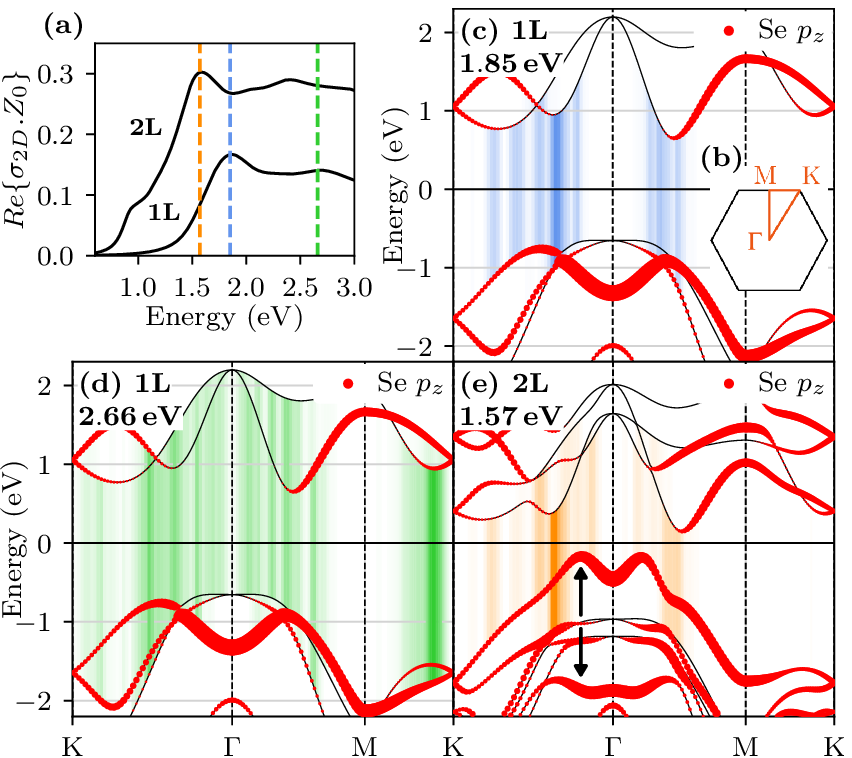}
\caption{Contributions to the optical absorption peaks. (a) DFT-computed optical absorption of 1L and 2L. (b) (Irreducible) Brillouin zone in (orange) black. (c\eb{}{-e}) Optical absorption contributions \eb{}{at given energies (dashed lines in (a)),} for 1L \eb{main peak, (d) second peak and (e) 2L main peak,}{and 2L absorption peaks,} decomposed on the \eb{$\rm{K\Gamma MK}$ band diagram}{electronic band structure plotted along the K-$\rm{\Gamma}$-M-K direction}. \eb{The color lines bind each eigenstate pair contributing to the optical absorption at the specified energy (dashed lines in (a)), the line opacity representing the contribution weight.}{Colored lines with varying opacity illustrate the relative contribution of pairs of eigenstates to the optical absorption.} \eb{In red is displayed t}{T}he selenium $p_z$ orbital component \eb{}{is displayed in red}, the line thickness giving its magnitude.}
  \label{fig:peaksContributions}
\end{figure}

\eb{The DFT-computed optical absorption (Fig. 2a) is decomposed on each available direct transition of the band diagram, for a given energy.}{Our numerical method allows to isolate the bands and k-wavevectors that contribute the most to the optical absorption for a given photon energy.}
The \eb{main}{$1.85\,\mathrm{eV}$ and $1.57\,\mathrm{eV}$} absorption peaks \eb{}{of 1L and 2L} arise from optical transitions located \eb{in}{around} the center of the irreducible Brillouin zone (Fig. \ref{fig:peaksContributions}a, b, c, e). \eb{For these transitions, the coupled bands are locally parallel,}{These transitions involve locally parallel valence and conduction bands,}  leading to local enhancements in the joint density of state (SM-\ref{SM-sec:absorptionPeaksJDOS}), and consequently \eb{in}{to} optical absorption peaks -- a situation known as the band nesting effect \cite{Carvalho2013BandNesting}. \eb{}{These peaks arise from optically active transitions from the symmetric selenium $p_z^{+}$ orbital to the platinum $d_{x^2-y^2/xy}$ orbital  (SM-\ref{SM-sec:orbitals}).}
\eb{In a similar fashion}{Similarly}, the \eb{second}{$2.66\,\mathrm{eV}$} absorption peak \eb{}{of 1L} originates from\eb{ band}{} nesting \eb{in}{of bands} between the M and K points (Fig. \ref{fig:peaksContributions}d) \eb{}{and involves} \eb{}{the Se $p_{x/y}^+$ and Pt $d_{zx/zy}$ orbitals}\eb{ coupling}{}.
\eb{The main absorption peak is mainly arising from optically active transitions from the symmetric selenium $p_z^{+}$ orbital to the platinum $d_{x^2-y^2/xy}$ orbital  (SM-4e).
The second peak principally originates from transitions based on the Se $p_{x/y}^+$ and Pt $d_{zx/zy}$ orbitals coupling.}{}

\eb{When stacking}{For} two \eb{}{stacked} monolayers, \eb{the interlayer coupling results in a lift of the bands degeneracy. T}{t}he interlayer hybridization is particularly strong for states composed of an important selenium $p_z$ orbital component (Fig. \ref{fig:peaksContributions}d, e\eb{, SM-4e}{}).
Consequently, an upshifted \eb{(}{}and \eb{}{a} downshifted\eb{)}{} valence band emerge\eb{s}{} in the bilayer band diagram (\eb{dark}{black} arrows Fig. \ref{fig:peaksContributions}e), causing a bandgap reduction.
\eb{We}{However, we} observe that \eb{this}{the upshifted band, which is the} upmost valence band\eb{}{,} is \eb{however }{}not involved in the main absorption peak. It\eb{ instead}{} plays \eb{}{instead} an important role in the low-energy optical absorption.


The slowly decaying near-infrared absorption tail of $\rm{PtSe_2}$ has been interpreted as originating from indirect interband absorption. We now provide the \eb{experimental }{}evidence that such a mechanism is not involved. Inspecting the computed optical absorption allows \eb{to }{}explain\eb{}{ing} the tail profile, only relying on direct optical transitions.
 
\begin{figure}[h]
\includegraphics[width=3.375in]{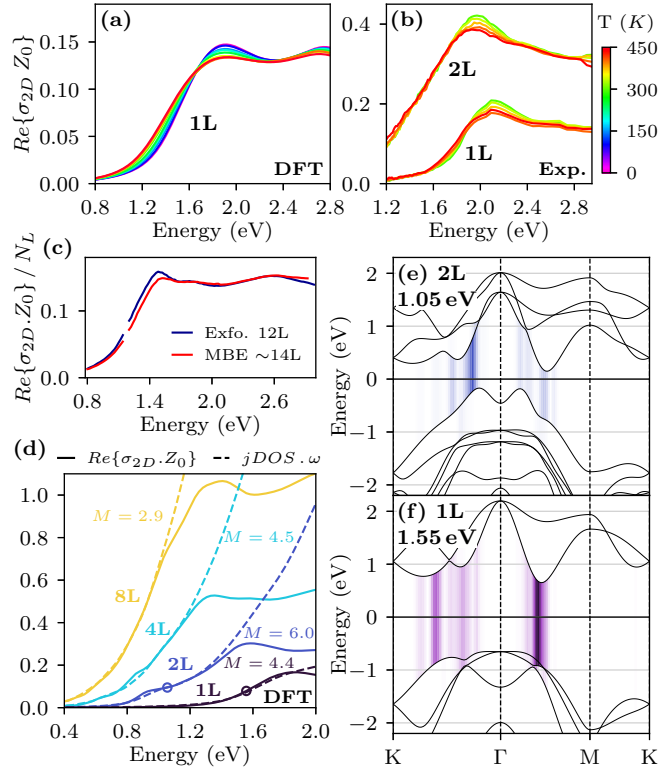}
\caption{Absorption tail origin. \eb{}{(a) Optical absorption $Re\{\sigma_{2D}.Z_0\}$ of 1L for $T =  0 - 450\,\rm{K}$ (computed) and of (b)} \eb{(a) Optical absorption $Re\{\sigma_{2D}.Z_0\}$ of} 1L and 2L for \eb{$290 - 440\,\rm{K}$}{$T =  290 - 440\,\rm{K}$ (experimental)}. (\eb{b}{c}) \eb{Rescaled exp.}{Experimental} real 2D optical conductivity of a 12L exfoliated flake and a $\sim$14L MBE-grow\eb{}{n} film \cite{Desgue2023MBE}\eb{}{, rescaled by the number of layers}. (\eb{c}{d}) Computed $Re\{\sigma_{2D}.Z_0\}$ (full line), and $jDOS\,.\,\omega$ (dashed line, $\omega$ is the optical frequency) for 1L, 2L, 4L and 8L -- rescaled by an \textit{ad hoc} dipolar matrix coefficient $M$ (expressed in Hartree atomic units). (\eb{d}{e}) 2L and (\eb{e}{f}) 1L contributions to the real 2D optical conductivity, at the energies circled in (\eb{c}{d}).
  \label{fig:absorptionTail}}
\end{figure}

\eb{In the case of indirect optical transitions, the momentum mismatch between the two coupled states can be provided either by the phonon bath or by defects in the material.}{}
\eb{}{To this end, we investigate the role of phonons}\eb{-assisted indirect transitions}{} \eb{}{on the optical absorption \cite{YuCardona2010ChapOptPtiesI} by using a} \eb{0-450K}{} \eb{}{temperature-dependent DFT simulation} \eb{}{that treats simultaneously direct and phonon-assisted indirect processes} \cite{Zacharias2016OneShotPRB, Zacharias2020SpecialDisplacement}. \eb{}{The absorption tail is mostly unaffected \footnote{A small energy shift is due to electronic band structure renormalization with temperature (see SM-\ref{SM-sec:DFT-temperature})} by a $0-450\,\mathrm{K}$ temperature increase}, \eb{as expected for}{indicating the dominance of direct transitions  (Fig. 3a)}.
\eb{}{This is confirmed experimentally by investigating 1L and 2L samples absorption while increasing the temperature from $290$ to $440\,\rm{K}$  (Fig. 3b)}. 

\eb{We observe that a 140 K increase does not broaden the absorption tail of the 2L sample, and marginally affects the one of the 1L sample (Fig. 3a, SM 3d), ruling out phonon-assisted indirect transitions mechanism. 
We also compare}{We also investigate the role of defects on optical absorption, by considering} an exfoliated flake and a thin film grown by molecular beam epitaxy \cite{Desgue2023MBE}. Both feature different Raman linewidths \cite{articlePtSe2Raman}, \eb{and should therefore have}{testifying of} different types and\eb{}{/or} densities of defects \cite{Pierce2018IrradGrMoS2Raman, Desgue2023MBE}. \eb{They however}{However, they} present almost identical absorption tails (Fig. \ref{fig:absorptionTail}\eb{b}{c})\eb{discarding}{}\eb{the hypothesis of a defect-assisted transition mechanism}{}. 


Since indirect transitions are experimentally ruled out, we come back to the direct optical transition interpretation guided by the insights provided by DFT.
We start by confronting the real \eb{}{part of the} 2D optical conductivity $\sigma_{2D}$ with the joint density of states $jDOS$ multiplied by frequency $\omega$\eb{(}{. As shown in }Fig. \ref{fig:absorptionTail}\eb{c}{d}\eb{).}{, they}
\eb{They }{}coincide over the absorption tail spectral range\eb{, which extends on a large energy range }{ (}$\sim 0.7\,\rm{eV}$\eb{}{)}, so that we can write $Re\{\sigma_{2D}\} \sim M\,.\,jDOS\,.\,\omega$. This \eb{form}{relation} is expected \eb{if the involved}{for} optical transitions \eb{feature}{featuring} a constant light-matter coupling, that we identify here as $M$\eb{ (SM-4bi)}{} -- suggesting that specific transitions, involving analogous pairs of states, are responsible for the tail absorption.
We indeed identify these transitions by inspecting the contributions to optical absorption of 1L and 2L (Fig. \ref{fig:absorptionTail}\eb{c}{e} and \eb{d}{f})\eb{. They}{: they} occur between the upmost $p_z^+$-based valence band and a conduction band bearing Pt $d_{x^2-y^2/xy}$ orbital component.\eb{ 
For the bilayer sample, a small shoulder appears at $0.9\,\rm{eV}$ which is not as prominent in the $jDOS$ (Fig. 3c). It indicates a change in the light-matter coupling $M$, most likely due to the onset of Se $p_{x/y}^+$ to Pt $d_{zx/zy}$ orbital coupling (SM-4bii
).}{}
Due to band degeneracy lift\eb{}{ing} when stacking individual layers, the number of optical transitions involved in the absorption tail increases with the layers count\eb{ (SM Fig. 11)}{}. Consequently, the monolayer sample exhibits resonant transitions only $0.3\,\rm{eV}$ below its \eb{main}{$1.85\,\mathrm{eV}$} peak, while the resonant transitions in 2L extend $0.7\,\rm{eV}$ below \eb{(Figs 2c, e, SM12, and 3e)}{(Figs \ref{fig:absorptionTail}\eb{c, e}{d, f})}.
This explains the steeper absorption decrease of 1L (Fig. \ref{fig:conductivitiesSchemes}d bottom), which in turn leads to a stronger sensitivity to temperature-induced broadening (Fig. \ref{fig:absorptionTail}a\eb{}{, b}).


\begin{figure}[h]
\includegraphics[width=3.375in]{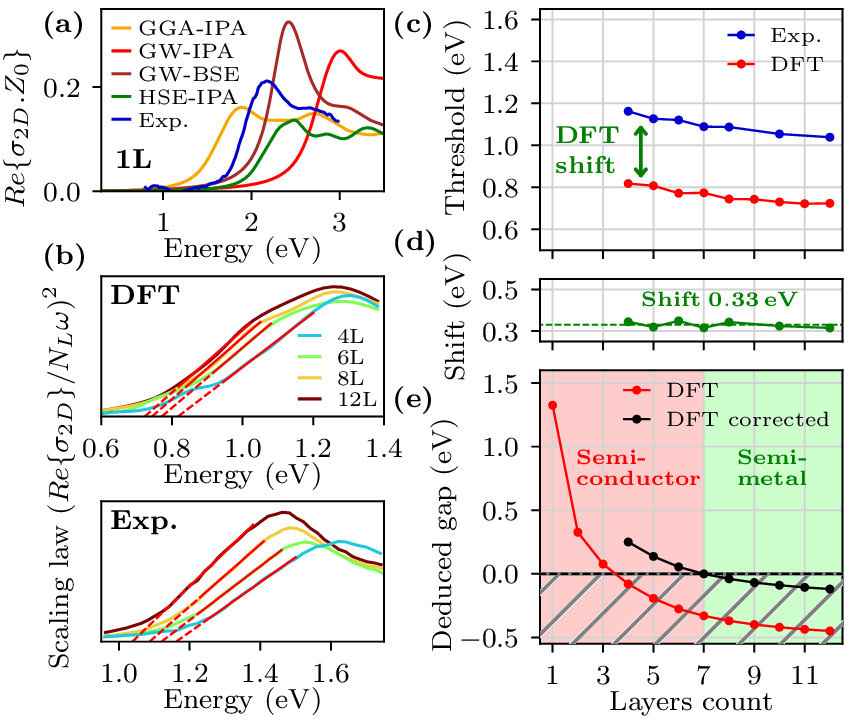}
\caption{DFT bandgap correction using the optical absorption tail. (a) Comparison of computed real 2D optical conductivities for monolayer $\rm{PtSe_2}$ using several DFT methods\eb{ (SM-4c)}{}. \eb{The dashed lines are eyeguides for the main peak position.}{}(b) Extraction of the tail threshold energy, for \eb{}{GGA} DFT-computed and \eb{exp.}{experimental} absorptions, using linear extrapolation from the $\left(Re\{\sigma_{2D}\}/N_L\omega\right)^2$ scaling law\eb{.}{ (}$\omega$ is the energy and $N_L$ the layer count\eb{}{)}. (c) \eb{Exp.}{Experimental} and theoretical threshold energ\eb{y}{ies} depending on \eb{}{the} layers count, and (d) the inferred energy shift. (e) DFT-computed bandgap \eb{}{(red)}, and its \eb{corrected }{}value \eb{(}{}upshifted by $0.33\,\rm{eV}$ \eb{}{(black}). \eb{}{The diagonal hatched area is the $0\,\mathrm{eV}$ gap region.}}
\label{fig:gapExtrapolation}
\end{figure}

While DFT is a tool providing valuable information about the \eb{strength and}{coupling of} electronic wavefunctions involved in optical transitions, its bandgaps estimates have limited accuracy.
\eb{We circumvent this problem by using the measured optical absorption to rescale the DFT calculations. In doing so, we determine the indirect bandgap energies and deduce the indirect semiconductor to semimetal transition.}{To illustrate this, we compare DFT-computed optical absorption using different DFT models (Fig. \ref{fig:gapExtrapolation}a), and observe a model-dependent energy shift of the absorption profile\eb{}{s}. 
This is caused by different accounts of many-body interactions, which appear as a rigid shift of the conduction bands with respect to the valence ones (SM-\ref{SM-sec:rigidShifthyp}) - a well-known limit of DFT methods \cite{Falco2013GwComputation}. This observation invites us to rescale the DFT calculations using the optical absorption measurements to determine the indirect bandgap energies and deduce the semiconductor-to-semimetal transition.}
\eb{DFT simulations are known to introduce a rigid energy shift of the conduction bands with respect to the valence bands due to misestimation of many-body interactions, leaving the bands profiles essentially unperturbed [44] (SM Fig. 13). We represent in Fig. 4a the optical absorption of 1L obtained with DFT models beyond GGA-IPA to capture many-body effects. However all the simulations present energy shifts with respect to the experimental results (Fig. 4a), motivating the correction of this shift. }{}\eb{To do so, we choose to}{More specifically, we} use the \eb{lowest energy}{NIR} optical signature: the absorption tail.
This \eb{feature}{latter} has been \eb{previously }{}used \eb{}{in prevous works} to extract $\rm{PtSe_2}$ bandgaps using Tauc plots. We now rely on the knowledge that this tail results from direct transitions, and derive a scaling law for the bulk limit\eb{.}{:}
close to the lowest energy direct transition $\omega_0$, the energy bands are locally quadratic in momentum and the light-matter coupling can be considered as constant, \eb{raising}{so that} $\left(Re\{\sigma_{2D}\}\,/\,\omega\right)^2 \sim \omega - \omega_0$ (SM-\ref{SM-sec:scalingLaw}).
Applying this law to the \eb{}{GGA} DFT-computed and experimental absorption tails, we extract threshold energies by extrapolating the linear slopes to the x-intercept, down to 4L thickness (Fig. \ref{fig:gapExtrapolation}b and c).
Note that as the material is thinned down, 2D electronic confinement leads to the discretization of the z-momentum. Hence, the scaling law only applies for thick enough materials (in practice 4L).
The inferred threshold values for experimental and simulated data feature a nearly constant $0.33\,\rm{eV}$ energy shift (Fig. \ref{fig:gapExtrapolation}d). We consequently upshift the DFT indirect bandgap predictions by this value \eb{}{(Fig. \ref{fig:gapExtrapolation}e)}.
With this correction, the transition from indirect semiconductor to semimetal\eb{lic}{} state occurs around 7 layers, in contrast with initial \textit{ab initio} calculations which predicted $\rm{PtSe_2}$ to be semiconducting up to 2-3 layers \cite{Villaos2019DFT, Kandemir2018RamanDFT}\eb{ (Fig. 4e)}{}.
Experimental electronic transport and scanning tunneling spectroscopy measurements reported transitions observed between 3 and 10 layers \cite{Ciarrocchi2018SmScAndrasKis, Yim2018TransportDC, Ansari2019ScSmGrown, Das2021ContactScSmPtSe2, Zhang2021STS, Li2021STM}. These estimations present a large spread, which could be explained by the presence of spurious conduction \eb{}{channels} due to defects or \eb{the}{a} lack of accuracy in the determination of the number of layers. In our work, by employing optical absorption spectroscopy we precisely establish the number of layers, and the use of an indirect method for determining the electronic bandgap makes it insensitive to the presence of defects.
A 7-layer semiconductor-to-semimetal transition opens the possibility to fine-tune the electronic bandgap down to very low values -- close to the transition, the bandgap shifts by only $\sim 50\,\rm{meV}\,/\,\mathrm{layer}$ -- of particular interest for middle and far infrared optoelectronics.

\begin{figure}[h]
\includegraphics[width=3.375in]{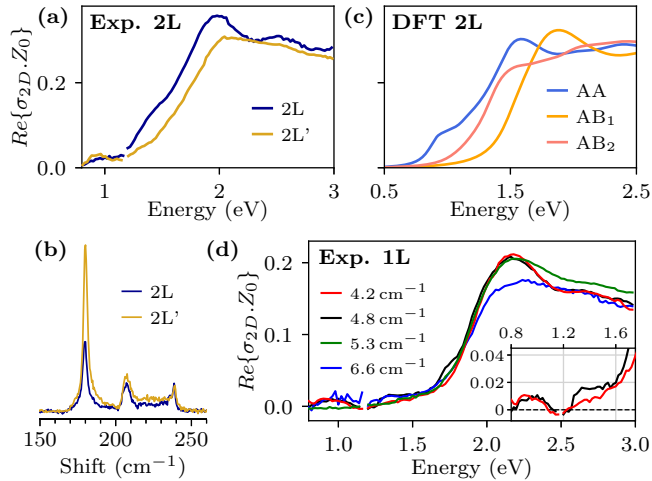}
\caption{Bilayer stacking and monolayer presumed bound exciton\eb{s}{}. (a) Experimental optical absorption of two bilayers\eb{}{, (b)} featuring different Raman signatures\eb{ (insets)}{}. (\eb{b}{c}) DFT-computed optical absorption of 2L with AA\eb{ or AB}{, $\mathrm{AB_1}$ or $\mathrm{AB_2}$} stackings (\eb{}{structure in} SM-\ref{SM-sec:ABStacking}). (\eb{c}{d}) Exp\eb{.}{erimental} monolayers optical absorption, labeled by their Raman $E_g$ mode linewidth \cite{articlePtSe2Raman}, and (inset) low-energy magnification.}
\label{fig:case1L2L}
\end{figure}

The optical absorption mechanism demonstrated in our study is supported by robust sample-to-sample reproducibility\eb{ (SM Fig. 7)}{}. Mono and bilayers are however notable exceptions (Fig. \ref{fig:case1L2L}a, \eb{c}{d})\eb{}{, which we further consider below}.

The nature of the interlayer coupling strongly affects 2D materials optical absorption \cite{Liang2022StackingMoS2}, and in particular \eb{those}{the one} of bilayer \eb{$\rm{1T-PtSe_2}$}{$\rm{PtSe_2}$}.
The Raman spectral signature allows to sort bilayers in two categories \cite{articlePtSe2Raman}, labeled 2L and 2L' (Fig. \ref{fig:case1L2L}\eb{a}{b}).
With respect to 2L samples, the 2L' \eb{ones }{}display a blue-shifted \eb{main peak}{optical absorption}, do not have an energy shoulder at $1.4\,\rm{eV}$, and exhibit a low intensity absorption peak at $1.0\,\rm{eV}$.
These three features are shared with monolayer  $\rm{PtSe_2}$ (Fig. \ref{fig:case1L2L}\eb{c}{d}), suggesting that the 2L' samples have a reduced interlayer coupling.
If AA stacking \eb{}{(1T phase)} is the most stable configuration of \eb{$\rm{1T-PtSe_2}$}{octahedral $\rm{PtSe_2}$}, AB stacking has been predicted to be stable as well \cite{Fang2019AA_AB_DFT, Kandemir2018RamanDFT, Kempt2022Stacking} and has been found to be a common defect of grown films \cite{Desgue2023MBE, Xu2021STEMstackAB, Ryu2019STEMstackAB}. Comparing the simulated optical absorption of AA and of two types of AB stacking (Fig. \ref{fig:case1L2L}\eb{b, SM-4d}{c}), we find that the AB stacked spectra present a characteristic blueshift, and do not \eb{have}{display} the $0.9\,\rm{eV}$ characteristic shoulder of the AA stacked spectrum.
The optical absorption of 2L supports therefore an AA\eb{ }{-}stacked configuration, while the one of 2L' is likely caused by AB stacking.

We now turn to the $\rm{PtSe_2}$ monolayers: the sample of worst crystalline quality, featuring a Raman $E_g$ mode linewidth above $6\,\rm{cm^{-1}}$ \cite{articlePtSe2Raman}, presents a reduced \eb{main}{$2.2\,\mathrm{eV}$} peak amplitude with respect to the best samples.
This is not surprising as this peak results from band nesting effect, a singularity of the band structure that can \eb{}{be }particularly \eb{be }{}altered by the presence of disorder -- even more since it is affected by many-body interactions (GW-BSE in Fig. \ref{fig:gapExtrapolation}a).

Beyond redistribution of oscillator strength caused by many-body effects, one can wonder if bound excitons might be observed in $\rm{PtSe_2}$. We \eb{}{have} perform\eb{}{ed} a photoluminescence experiment at room temperature within the visible range (SM-\ref{SM-sec:PL}). The measured signal is very weak, and no distinctive excitonic signatures can be identified.
Inspecting the optical absorption of monolayer $\rm{PtSe_2}$, an unexpected feature is nonetheless observed at lower energy: the high-quality samples, of Raman mode linewidth below $5\,\rm{cm^{-1}}$, exhibit a $1\,\%$ absorption peak at $1.0\,\rm{eV}$ (Fig. \ref{fig:case1L2L}\eb{c, SM-3e}{d}).
This peak is located below the absorption tail, in the optical gap, and might well originate from a bound exciton. Future work is needed \eb{with higher-quality encapsulated samples }{}at low temperature to investigate further its origin.


In conclusion, we have shown that the NIR-vis optical absorption -- peaks and low-energy tail -- of indirect bandgap thin layered $\rm{PtSe_2}$ arises from single-electron direct optical transitions. As a consequence, we could infer its semiconductor\eb{ }{-}to\eb{ }{-}semimetal transition at 7 layers. This direct and tunable optical absorption paves the way to efficient and ultrafast optoelectronics in the telecom band. Furthermore, the now available high-quality exfoliated $\rm{PtSe_2}$ down to the monolayer allows \eb{to }{}explor\eb{e}{ing} many-body effects, such as a potential monolayer exciton at $1.3\,\rm{\mu m}$.\\


We thank Y. Chassagneux and F. Rapisarda for experimental help.
The authors acknowledge the financial support from the European Union’s Horizon 2020 program under grant agreement no. 785219, no.  881603 (Core2 and 3 Graphene Flagship) and no. 964735 (FET-OPEN EXTREME-IR), as well as from ANR-2018-CE08-018-05 (BIRDS), ANR-2021-CE24-0025 (ELuSeM), ANR-20-CE09-0026 (2DonDemand), and Thales Systèmes Aéroportés (CIFRE grant No. 2016/1294).
This work was granted access to the HPC resources of MesoPSL financed by the Region Ile de France and the project EquipMeso (reference ANR-10-EQPX-29-01) of the program Investissements d'Avenir supervised by the Agence Nationale pour la Recherche.
Authors contributions are detailed in SM-\ref{SM-sec:authorsContributions}.
Data are publicly available at
\href{https://doi.org/10.5281/zenodo.10138281}{10.5281/zenodo.10138281}.


\bibliography{ref}

\makeatletter\@input{xx.tex}\makeatother

\end{document}


\title{The optical absorption in indirect semiconductor to semimetal \texorpdfstring{PtSe\textsubscript{2}}{PtSe2} arises from direct transitions -
Supplemental Material}

\author{Marin Tharrault}
\affiliation{\ens}
\author{Sabrine Ayari}
\affiliation{\ens}
\author{Mehdi Arfaoui}
\affiliation{\fst}
\author{Eva Desgué}
\affiliation{\trt}
\author{Romaric Le Goff}
\affiliation{\ens}
\author{Pascal Morfin}
\affiliation{\ens}
\author{José Palomo}
\affiliation{\ens}
\author{Michael Rosticher}
\affiliation{\ens}
\author{Sihem Jaziri}
\affiliation{\fst}
\affiliation{\fsb}
\author{Bernard Plaçais}
\affiliation{\ens}
\author{Pierre Legagneux}
\affiliation{\trt}
\author{Francesca Carosella}
\affiliation{\ens}
\author{Christophe Voisin}
\affiliation{\ens}
\author{Robson Ferreira}
\affiliation{\ens}
\author{Emmanuel Baudin}
\affiliation{\ens}
\affiliation{\iuf}

\maketitle

\tableofcontents

\section{Au-assisted mechanical exfoliation} \label{SM-sec:AuAssistExfo}

\begin{figure}[h!]
\begin{center}
\includegraphics[width=4in]{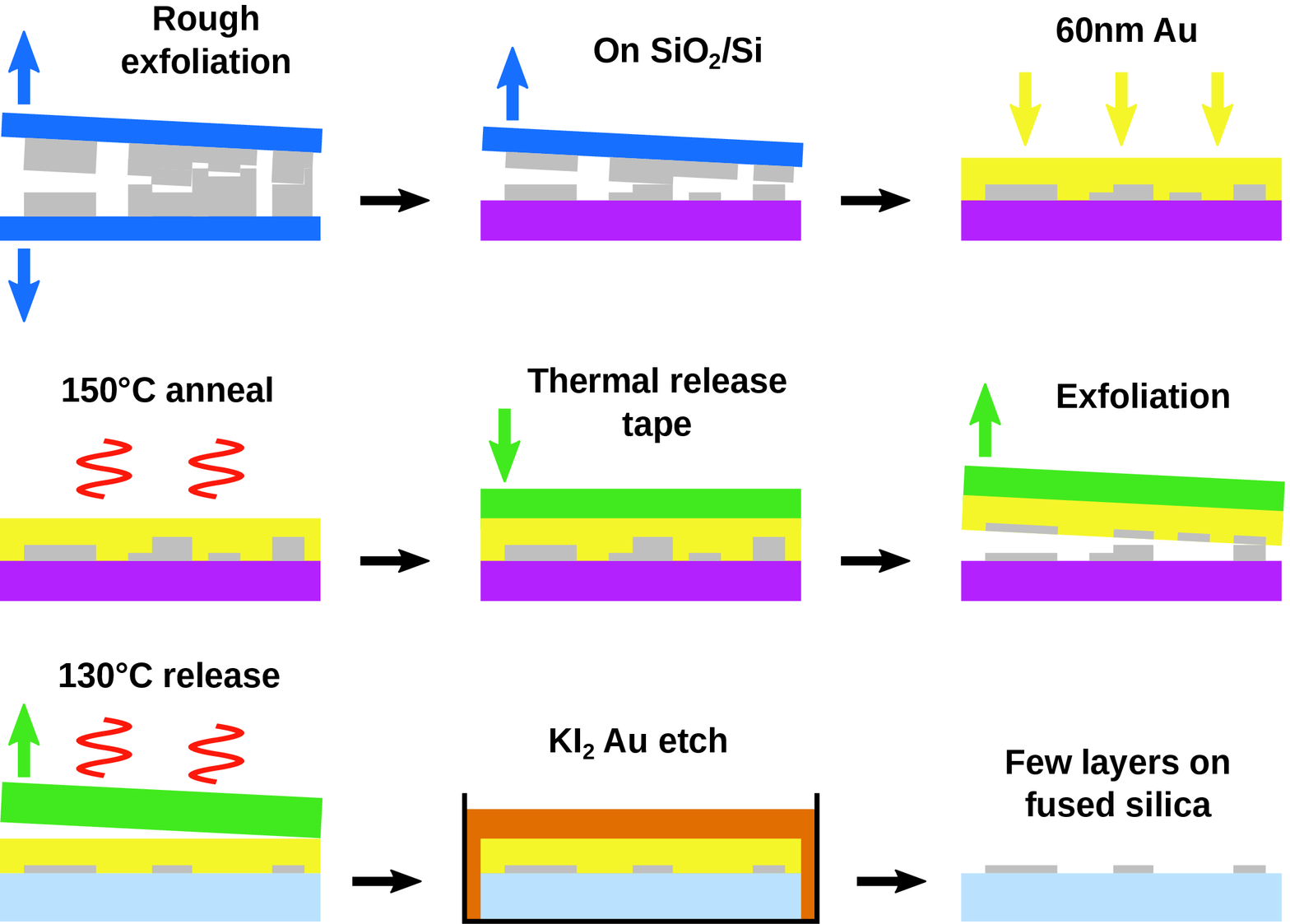}
\end{center}
\vspace{-2ex}
\caption{Au-assisted exfoliation protocol.}
\label{SM-fig:auExfoliation}
\end{figure}

The process used to mechanically exfoliate few-layers $\rm{PtSe_2}$ flakes is adapted from the one developed by Desdai \textit{et al.} \cite{Desdai2016AuExfo}, and is schematized in Fig. \ref{SM-fig:auExfoliation}.
A first coarse exfoliation of a bulk crystal grown by CVT (HQ graphene) is performed with standard blue tape onto a clean $\rm{SiO_2}$/$\rm{Si}$ substrate.
The sample is cleaned in acetone/IPA and a $60\,\rm{nm}$ Au layer is deposited by Joule evaporation. It is followed by a $150\rm{^\circ C}$ anneal in order to create differential dilation between Au and the $\rm{PtSe_2}$ stack \cite{Heyl2020AuExfoAnneal} thereby detaching few layers of crystal. Finally the Au layer is peeled off using thermal release tape and freed on a fused silica substrate, alongside with few layers $\rm{PtSe_2}$ flakes. The glue residues on the top layer of Au are removed using acetone/IPA from and the Au is etched with a solution of $\rm{KI_2}$ (4 $\rm{KI}$ : 1 $\rm{I_2}$ : 400 $\rm{H_2O}$, about $100 - 200\,\rm{nm/min}$).

The narrow Raman spectral peaks recorded, compared to those originating from $\rm{PtSe_2}$ films obtained with CVD, MBE or mechanical exfoliation in other works \cite{articlePtSe2Raman}, is an assessment of the structural quality of the obtained flakes, apparently unperturbed by the Au-assisted exfoliation technique.

\section{Atomic force microscopy - thickness determination} \label{SM-sec:AFM}

\begin{figure}[h!]
\begin{center}
\includegraphics[width=4.7in]{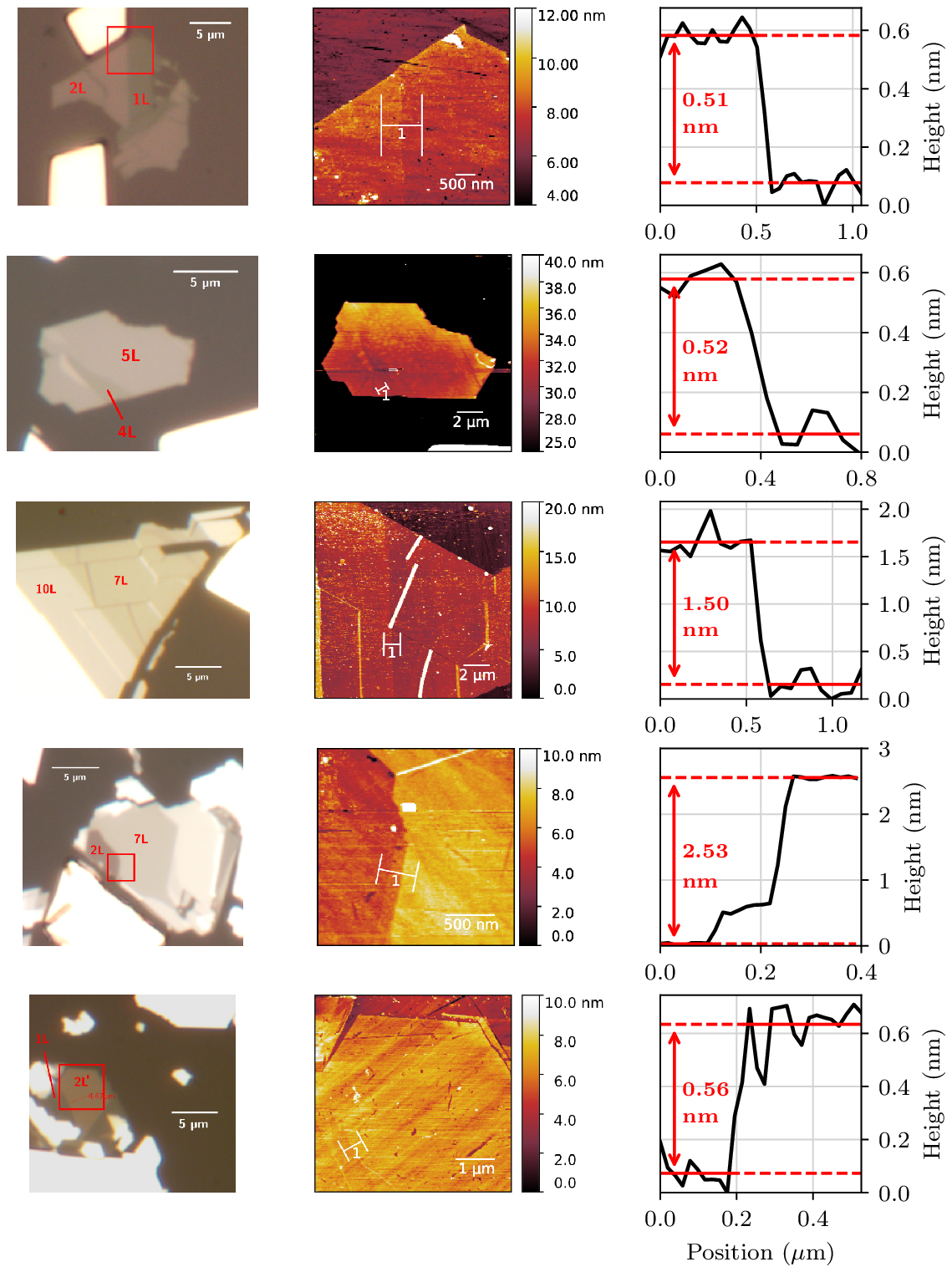}
\end{center}
\vspace{-5.5ex}
\caption{$\rm{PtSe_2}$ flakes AFM step heights. $\rm{1^{st}}$ column: pictures of $\rm{PtSe_2}$ flakes, each area labeled with its layers count. When needed, the area studied is indicated as a red square. Very bright areas are gold pads used to normalize the reflection, these pads are not present in the AFM scan. $\rm{2^{nd}}$ column: AFM height mappings. Cuts are indicated with "1" labels.  $\rm{3^{rd}}$ column: Height profiles along the cut. Using a layer thickness of $0.5\,\rm{nm}$, we deduce a step height of 1, 1, 3 and 5 layers from top to bottom flakes, in agreement with the optical absorption identification.\label{SM-fig:afmHeights}}
\end{figure}

Identification of the layer number of the exfoliated flakes is not obvious on such a new material as $\rm{PtSe_2}$. The absorption in the green-blue range $2.5-3.0\,\rm{eV}$ appears to be the best means for such a characterization, since it evolves linearly with the flake thickness. In order to provide consistent proof of it, the deduced thickness  has been crossed-checked with Atomic Force Measurement (AFM), using a Dimension Edge Bruker system.

It is well known that measuring a flake to substrate step height is not a reliable method to determine the flake's absolute number of layers: an interface layer of water is often present in between the flake and the substrate, which can reach $\sim 1\,\rm{nm}$. Yet, the step height between flakes of different thicknesses reliably establishes their layer count difference. As depicted in Fig. \ref{SM-fig:afmHeights}, the layer count identified with absorption spectroscopy is in agreement with the measured step heights. The individual layer thickness deduced from these measurements is $0.5\,\rm{nm}$, in agreement with the one computed by DFT (section \ref{SM-sec:DFT}) and measured by x-ray diffraction by the crystal supplier on $\mathrm{PtSe_2}$ powder \footnote{On the XRD measurement of reference \cite{HQGraphenePtSe2}, the first diffraction peak being at  $2\theta=17.51\mathrm{^\circ}$ for $\lambda =1.541 \AA$, one can infer the c-axis parameter as  $\lambda/(2 \sin(\theta))=0.506\,\mathrm{nm}$  }.

Raman spectroscopy is used as a complementary tool to identify between the flakes the ones of identical signature -- and therefore identical number of layers -- e.g., to assess that the bilayers of top and bottom rows in Fig.  \ref{SM-fig:afmHeights} are of identical thicknesses.

\section{Micro-absorption spectroscopy} \label{SM-sec:microAbsorption}

In this section we present the operational functioning and details of the micro-absorption experiment.

\begin{figure}[h!]
\begin{center}
\includegraphics[width=5.5in]{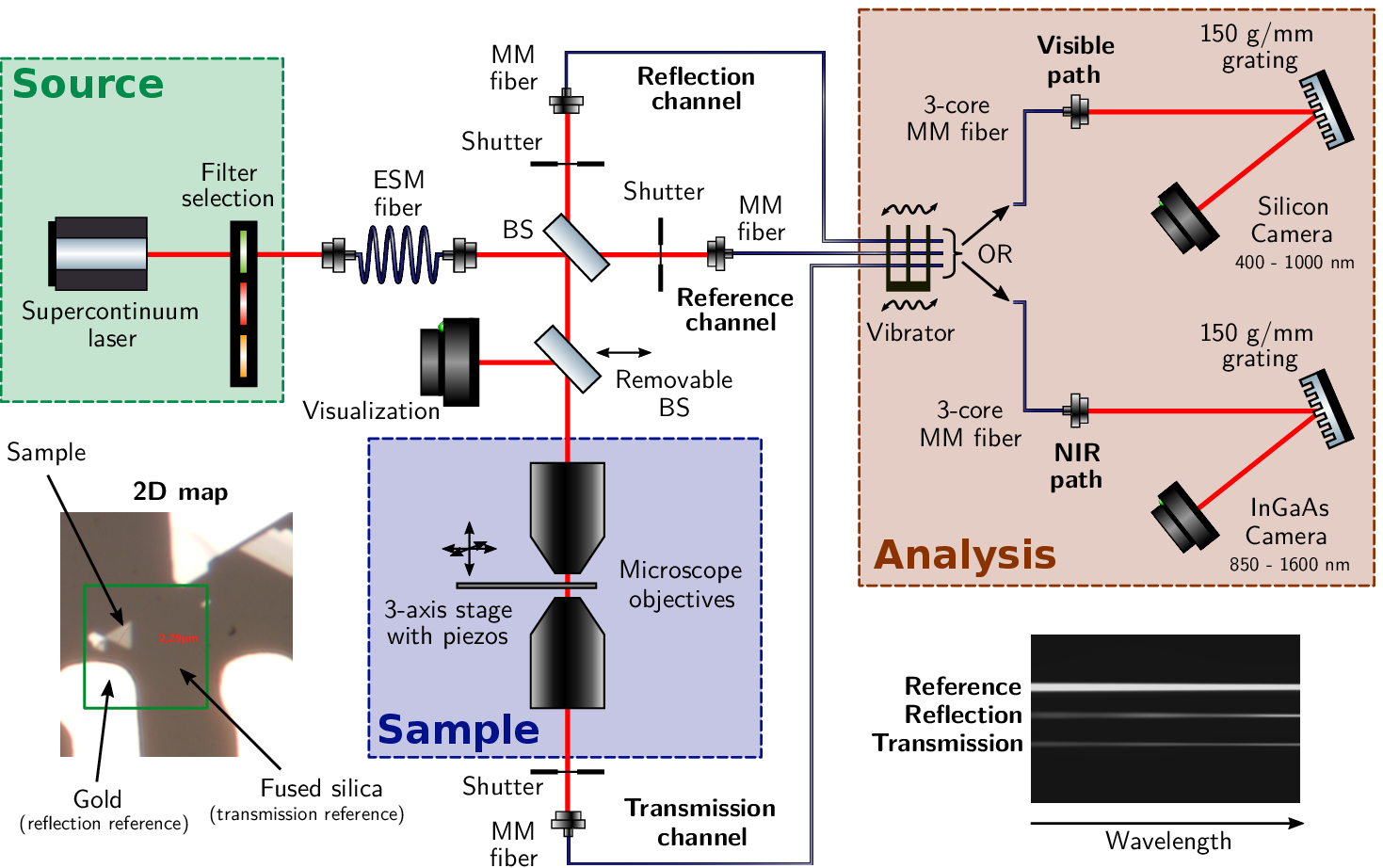}
\end{center}
\vspace{-3ex}
\caption{
Schematic of the micro-absorption experiment. \textbf{Source}: polychromatic light is emitted by a super-continuum laser and transmitted using achromatic optical elements. \textbf{Sample}: a pair of microscope objectives focus the light on the sample and collect its reflection and transmission. \textbf{Analysis}: spectrometers in the visible or in the infra-red record reflection, transmission and reference channels spectra. 2D mappings allow measuring a specific flake in addition to the highly reflective gold and transmissive fused silica reference.
  \label{SM-fig:schemeMicroAbsorption}}
\end{figure}

\subsection{Working principle}
Polychromatic light ($400 - 2200\,\rm{nm}$) is produced by a supercontinuum laser, filtered to avoid second order diffraction on the spectrometers gratings, and mode-cleaned using an endlessly single-mode fiber, as depicted on Fig. \ref{SM-fig:schemeMicroAbsorption}.
This light is focused with high numerical aperture apochromatic microscope objectives (one for the visible, one for the NIR) onto the sample (typical power $30\,\rm{\mu W}$ over a bandwidth of $300\,\rm{nm}$). The sample position is scanned using 3-axes piezoelectric positioners.
Reflected and transmitted light as well as a power reference are collected by multimode fibers and transmitted to visible or NIR spectrometers. The fibers are coiled around a vibrating device to time-average spurious speckles, and then combined into a 3-core multi-mode fiber. The output light from the fiber is focused on the input slit of the spectrometer. Gratings of $150\,\rm{g/mm}$ are used, enabling nanometric resolution as well as a large spectral range. The full visible / NIR spectra are reconstituted by combining several acquisitions.
Such spectra are collected while 2D mappings are performed, including the region of interest and absolute references for the reflection and transmission: highly reflective evaporated $120\,\rm{nm}$ gold surfaces and the transmissive fused silica substrate, used to normalize \textit{in situ} the reflection and the transmission respectively (reference data from \cite{McPeak2015AuRefractiveIndex, Malitson1965Sio2RefractiveIndex}).

The supercontinuum laser presents a rather flat spectral power density, at the notable exception of the wavelengths close to its pump power at $1064\,\rm{nm}$ where it is an order of magnitude larger. We therefore use a notch filter to suppress this spectral range, which appears in the data as a gap, located at $1.2\,\rm{eV}$ (main text Fig. \ref{fig:conductivitiesSchemes}d).

\subsection{Corrections and performances}
\subsubsection{Focus corrections and spatial resolution}
While capturing 2D mappings, a focus plane is used to take into account small sample misorientation. 
Microscope objectives focal shifts are as well corrected on the full visible to NIR range. Additionally, a refocalizing system is periodically scanning for the optimal focus to compensate long-term thermal shifts.

\begin{figure}[h!]
\begin{center}
\includegraphics[width=4.5in]{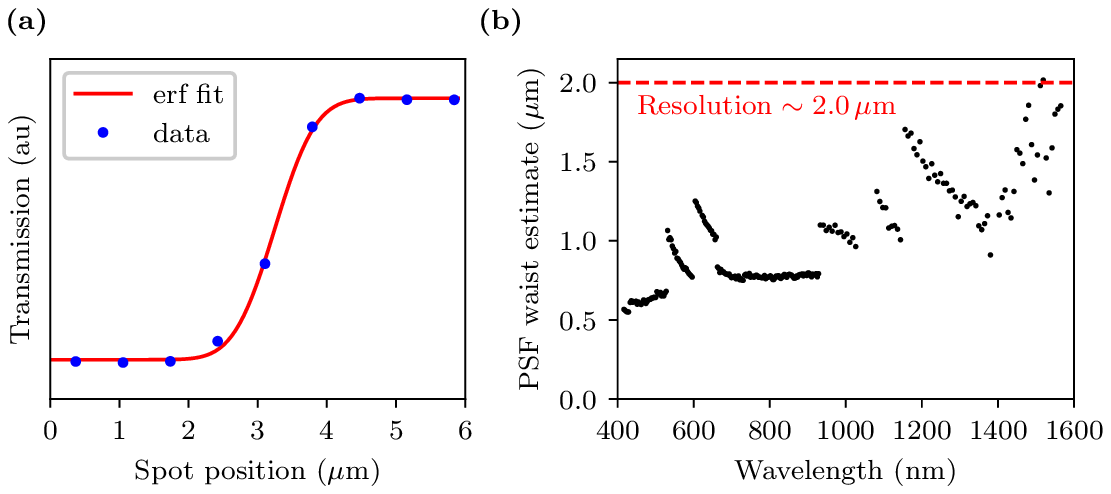}
\end{center}
\vspace{-4.5ex}
\caption{Spatial resolution measured with the knife-edge method. (a) Transmitted light while moving the spot from an exfoliated flake to the substrate. The data points are fitted with an error function, modeling the convolution of the PSF with a step-function. (b) Spectral point spread function (PSF) waist estimate from the fit, taken as twice the Gaussian standard deviation. Its maximum value provides the setup spatial resolution, that is about of  $2.0\,\rm{\mu m}$. The discontinuous nature of the plot is originating from the adjustment of the microscope objective focus while changing of wavelength range. The data used for the knife-edge method originate from a typical hyperspectral mapping.
\label{SM-fig:setupSpatialResolution}}
\end{figure}
Adding to these precautions a careful alignment of the microscope objective leads to a micrometric spatial resolution over the whole spectral range.  The spatial resolution can be measured using the knife edge method Fig. \ref{SM-fig:setupSpatialResolution}, giving a value of $2.0\,\rm{\mu m}$.

\subsubsection{Finite numerical aperture effects}

\begin{figure}[h!]
\begin{center}
\includegraphics[width=7in]{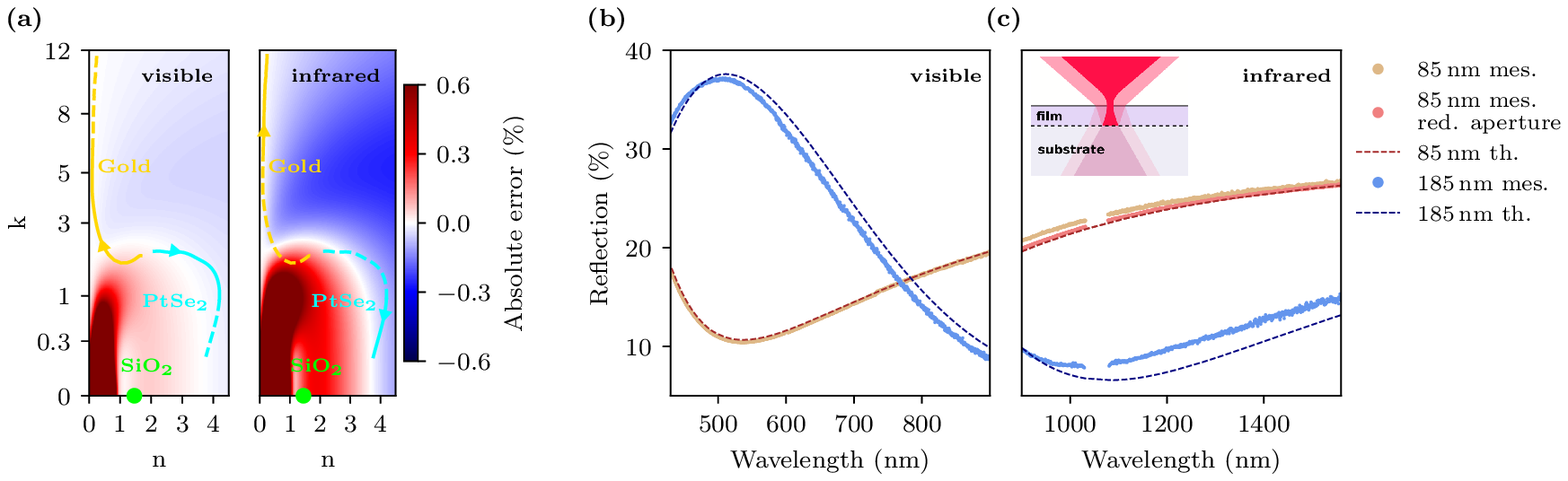}
\end{center}
\vspace{-5ex}
\caption{Finite numerical aperture (NA) effects on the measurement accuracy. (a) Computation of the absolute error made on the reflection by neglecting finite numerical aperture effects, as a function of the real and imaginary parts of the substrate's refractive index $n+\mathbf{i}k$. It is displayed as a colormaps for the objectives used in the visible ($\rm{NA} = 0.5$, left) and in the NIR ($\rm{NA} = 0.7$, right). The green dots, solid yellow and blue lines display the refractive indices of $\rm{SiO_2}$, gold and $\rm{PtSe_2}$ respectively, the arrow pointing towards increasing wavelengths. The error on the transmission at the interface is the opposite of the error on the reflection pictured in panel (a). (b) Measured (dotted lines) and computed (dashed lines) reflection for an $85\,\rm{nm}$ (brown) and $185\,\rm{nm}$ (blue) thick $\rm{SiO_2}$ layer on top of a $\rm{Si}$ substrate, on the visible range. (c) Same measurement performed on the NIR range. The red dots are measured on $85\,\rm{nm}$ $\rm{SiO_2}$/$\rm{Si}$ while reducing the microscope objective aperture by a factor of 2. (inset) Scheme picturing the internal reflections in the $\rm{SiO_2}$ slab. The $\rm{SiO_2}$/$\rm{Si}$ stepwafer has been calibrated by ellipsometry. The Refractive indexes for the $\rm{SiO_2}$, gold, $\rm{PtSe_2}$ and $\rm{Si}$ are extracted from \cite{McPeak2015AuRefractiveIndex, Malitson1965Sio2RefractiveIndex, Ermolaev2021EllipsOscillator, Green2008SiRefractiveIndex}.
\label{SM-fig:finiteNaAccuracy}}
\end{figure}

In this section, we investigate errors that may arise when using a focused light beam in our experiment. The thin film model utilized to retrieve intrinsic optical properties assumes that the incoming beam is at normal incidence with respect to the surface. Neglecting these errors may not seem obvious considering the high numerical apertures of the microscope objectives. We estimate consequently the error made on the reflection or transmission of $\rm{PtSe_2}$, gold or fused silica interface (by modelling a uniform illumination of the microscope aperture). We study separately visible and NIR ranges (which rely on different microscope objectives), and display the estimates using color maps (Fig. \ref{SM-fig:finiteNaAccuracy}a).
We find an error that reaches in absolute value at most $0.3\,\rm{\%}$, which is below the setup accuracy.

We moreover impose that the reflections interfering within the film perfectly overlap with each other -- as opposed to a situation where a thicker film would have internal reflections whose spot diameters present increasing diameters, as pictured in the inset of Fig. \ref{SM-fig:finiteNaAccuracy}c. This condition enables exact modelling of the interference phenomenon and eventually permits precise determination of the optical indices.
To quantify this phenomenon, we compare the reflection measured from an ellipsometry calibrated $\rm{SiO_2}$/$\rm{Si}$ step-wafer with its theoretically expected value.
In the visible -- using the appropriate microscope objective -- the measured reflectance features a shift compared to the predicted one for an $185\,\rm{nm}$ thick $\rm{SiO_2}$ layer while it shows good agreement for an $85\,\rm{nm}$ thick $\rm{SiO_2}$ layer (Fig. \ref{SM-fig:finiteNaAccuracy}b).
The effect is more pronounced in the infrared, since both stepwafers reflections present a shift with the expected values -- because of the higher numerical aperture of the objective. This effect can be however suppressed by reducing the objective aperture diameter, enabling to recover a fair agreement between measured and predicted values.
Accurate measurements of reflection (or transmission) therefore require the film under study to be sufficiently thin, typically below $e_{\rm{SiO_2}} \sim 40\,\rm{nm}$ which translates in terms of $\rm{PtSe_2}$ thickness as:
\begin{equation}
    e_{\rm{PtSe_2}} < \frac{\delta}{n_{\rm{PtSe_2}}} \sim \frac{n_{\rm{SiO_2}}}{n_{\rm{PtSe_2}}} e_{\rm{SiO_2}} \sim 15\,\rm{nm}
\end{equation}
In this work we study at most 12L of $\rm{PtSe_2}$, which represents a thickness of about $6\,\rm{nm}$.

\subsubsection{Aberrations correction and accuracy}

The roughness of the $150\,\rm{g/mm}$ grating of the visible spectrometer is responsible for light scattering out of the first diffraction order. This aberration is tackled by independent channel acquisitions and background fitting. The Silicon and InGaAs cameras are moreover carefully operated in their linearity ranges.
Under these conditions, the acquired spectra feature a typical $0.5\,\%$ accuracy as it can be inferred from the $\rm{SiO_2}$/$\rm{Si}$ step-wafer reflection measurement (for $85\,\rm{nm}$ $\rm{SiO_2}$ in the visible range and $185\,\rm{nm}$ $\rm{SiO_2}$ with a reduced aperture in the NIR range, Fig. \ref{SM-fig:finiteNaAccuracy}b and c).

However the use of the gold metallic surface as a reflection reference is only applicable beyond the $500\,\rm{nm}$ gold 5d to 6s transition. Indeed this resonance enables the existence of plasmonic modes, which are strongly sensitive to the surface roughness, and in turn, impact the reflectivity \cite{Kolwas2020AuPlasmon, McPeak2015AuRefractiveIndex}. In order to correctly normalize data in this range, we rather use the fused silica as a reference for reflection. It is however poorly reflective, with a typical $4\,\%$ reflectance, and systematic errors made on this value (as in Fig. \ref{SM-fig:finiteNaAccuracy}a) are amplified while normalizing for the resulting reflectance.
To solve this issue, we offset the data on the $400-600\,\rm{nm}$ range where fused silica is used as reference to match the one on the $> 600\,\rm{nm}$ range where the reference is the gold surface (this offset reaches typically $0.5\,\%$).

\subsection{Optical conductivity calculation} \label{SM-sec:optConductCalculation}
\begin{figure}[h!]
\begin{center}
\includegraphics[height=3.5in]{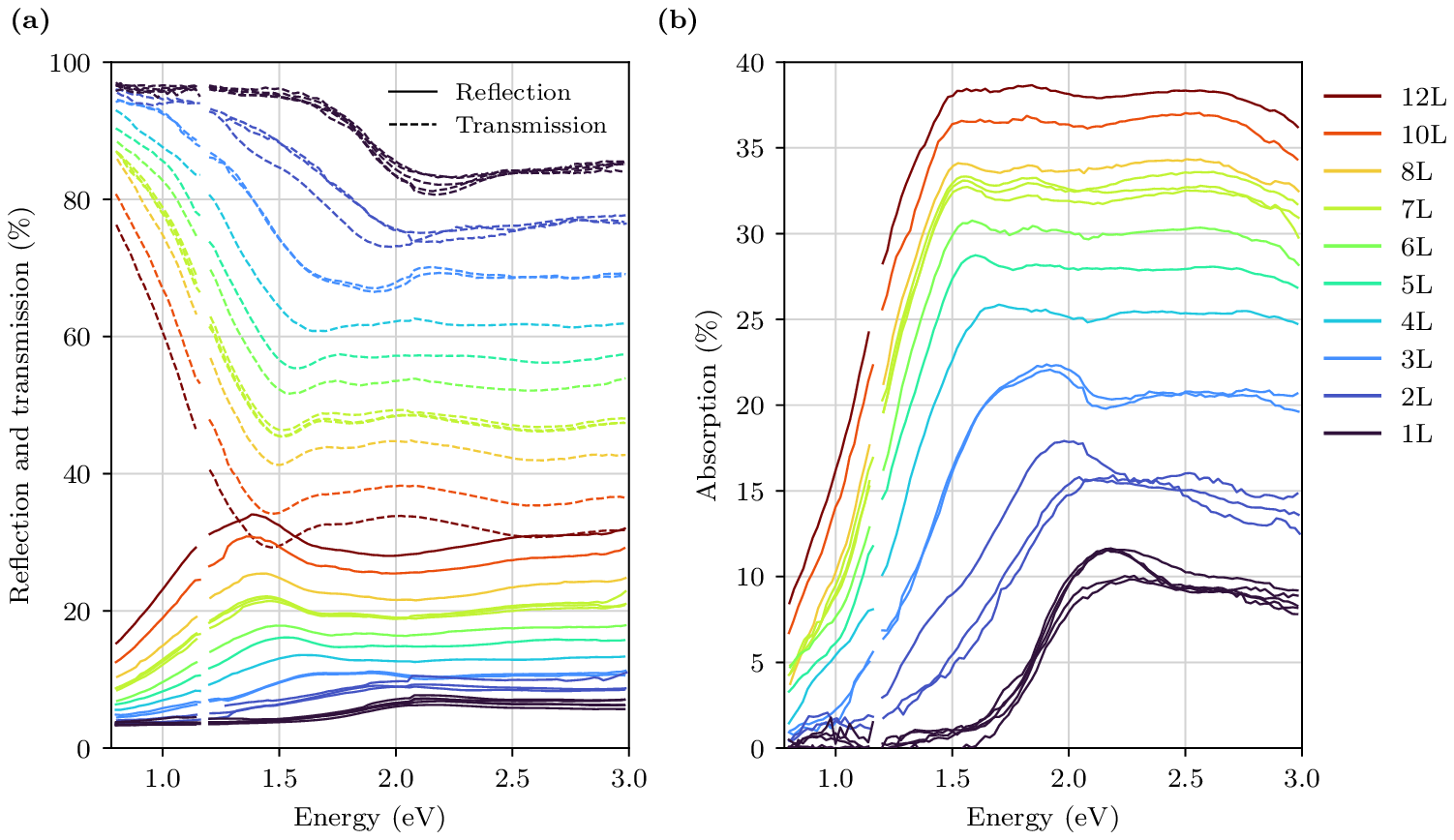}
\end{center}
\vspace{-4.5ex}
\caption{Absolute optical spectra of $\rm{PtSe_2}$ flakes on a fused silica substrate. (a) Reflection, transmission and (b) absorption channels.}
\label{SM-fig:RTAFull}
\end{figure}

\begin{figure}[h!]
\begin{center}
\includegraphics[height=4.5in]{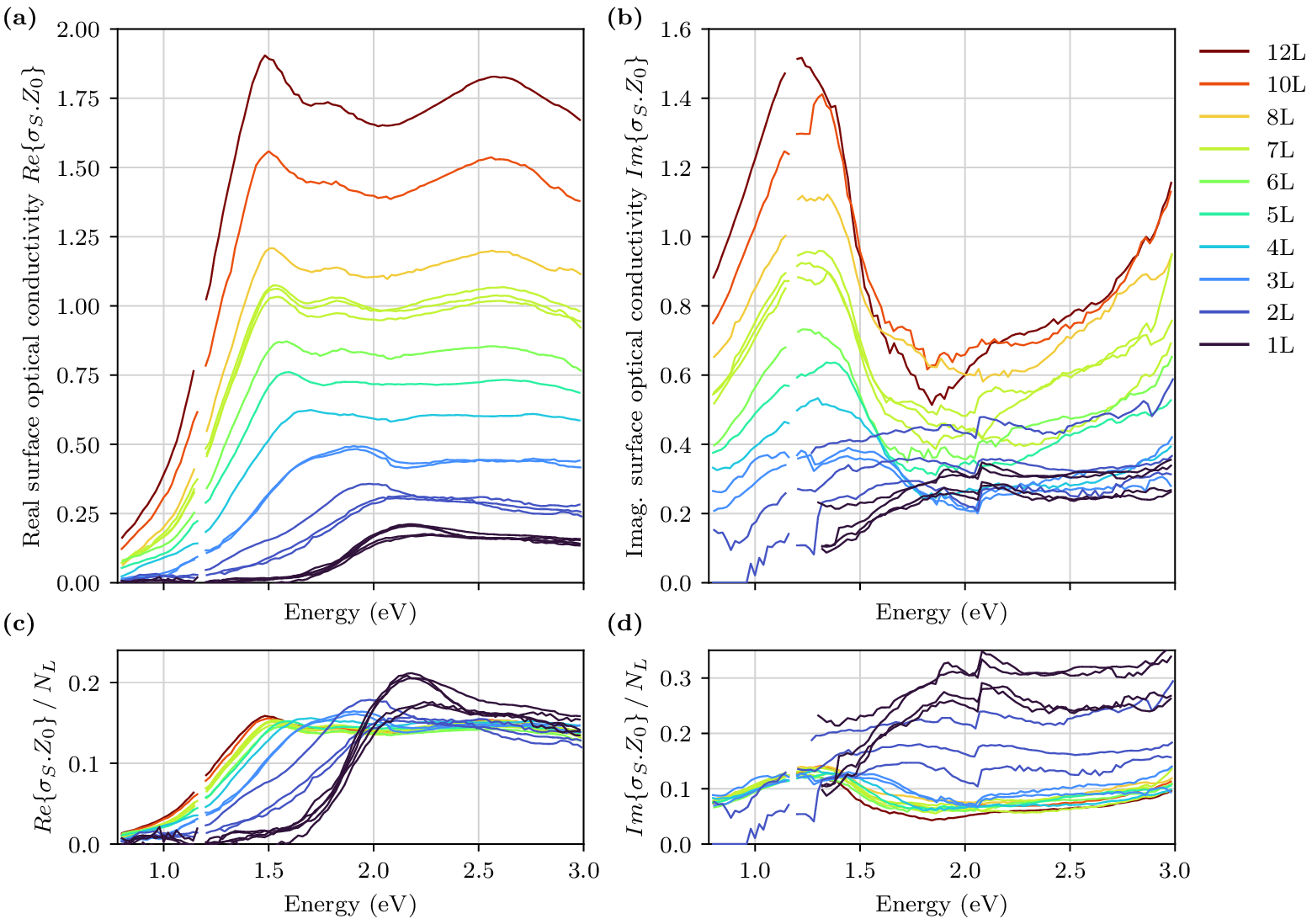}
\end{center}
\vspace{-5ex}
\caption{Measured 2D optical conductivity. (a) Real, and (b) imaginary parts of the optical conductivity, and their respective values divided by the number of layers (c and d). Data of $Im\{\sigma_{2D}\}$ is discarded for 1L low absorption values.}
\label{SM-fig:conductivityRealImagExperimental}
\end{figure}

From the spectral data of absolute reflection $R$ and transmission $T$ (Fig. \ref{SM-fig:RTAFull}), we can infer the 2D complex optical conductivity $\sigma_{2D}$ of the films by modelling their reflection and transmission at the air / substrate interface \cite{LiHeinz2018OptConduct}:

\begin{equation}
    R = \left|\frac{1 - n_S - Z_0\,\sigma_{2D}}{1 + n_S + Z_0\,\sigma_{2D}}\right|^2 \quad ; \quad T = Re\{n_S\}\left|\frac{2}{1 + n_S + Z_0\,\sigma_{2D}}\right|^2
\end{equation}

where $n_S$ is the substrate optical index and $Z_0 = 377\,\Omega$ is the impedance of the vacuum. The resulting spectra for 1 to 12 layers are displayed in Fig. \ref{SM-fig:conductivityRealImagExperimental}, binned using a $20\,\rm{meV}$ span.

As detailed by Li and Heinz \cite{LiHeinz2018OptConduct}, the differential reflection, differential transmission and absorption scale -- up to a constant factor -- with the real part of the optical conductivity $Re\{\sigma_{2D}\}$ (for a dielectric substrate and a film sufficiently thin).
Therefore, the method we use derives unambiguously the real part of the optical conductivity $Re\{\sigma_{2D}\}$. It also allows to derive the imaginary part of the optical conductivity $Im\{\sigma_{2D}\}$, but for a non-vanishing optical absorption.
On the other side, when using DFT, $Im\{\sigma_{2D}^{th}\}$ is inferred by Kramers-Kronig inversion of $Re\{\sigma_{2D}^{th}\}$ (which involves a long-range integral), and is therefore very sensitive to oscillator strength misestimations.
Consequently the data of $Im\{\sigma_{2D}\}$ are not commented in our work.

The complex optical conductivity is the adequate quantity to quantify a thin film optical response, as the resulting quantity is independent of the slab thickness.
Considering the complex optical conductivity instead of the bulk dielectric permittivity is permitted under the assumption of a negligible phase shift in the thin slab.
This can be done under the condition $\left|Z_0 \sigma_{2D}\right| \ll|n|$ \cite{LiHeinz2018OptConduct}. In our case, $\left|n_{\rm{PtSe_2}}\right| \sim 4$ (Fig. \ref{SM-fig:finiteNaAccuracy}a), while $\left|Z_0 \sigma_{2D}\right|$ reaches at most 2 for 12L stacking. This condition is therefore not strictly respected for the thick flakes. However, the excellent agreement between the normalized real conductivities of 10L and 12L (Fig. \ref{SM-fig:conductivityRealImagExperimental}c) gives confidence in the fair accuracy of the procedure.

\subsection{Temperature-dependent differential reflection} \label{SM-sec:thermalDiffRelf}

\begin{figure}[h!]
\begin{center}
\includegraphics[width=5.5in]{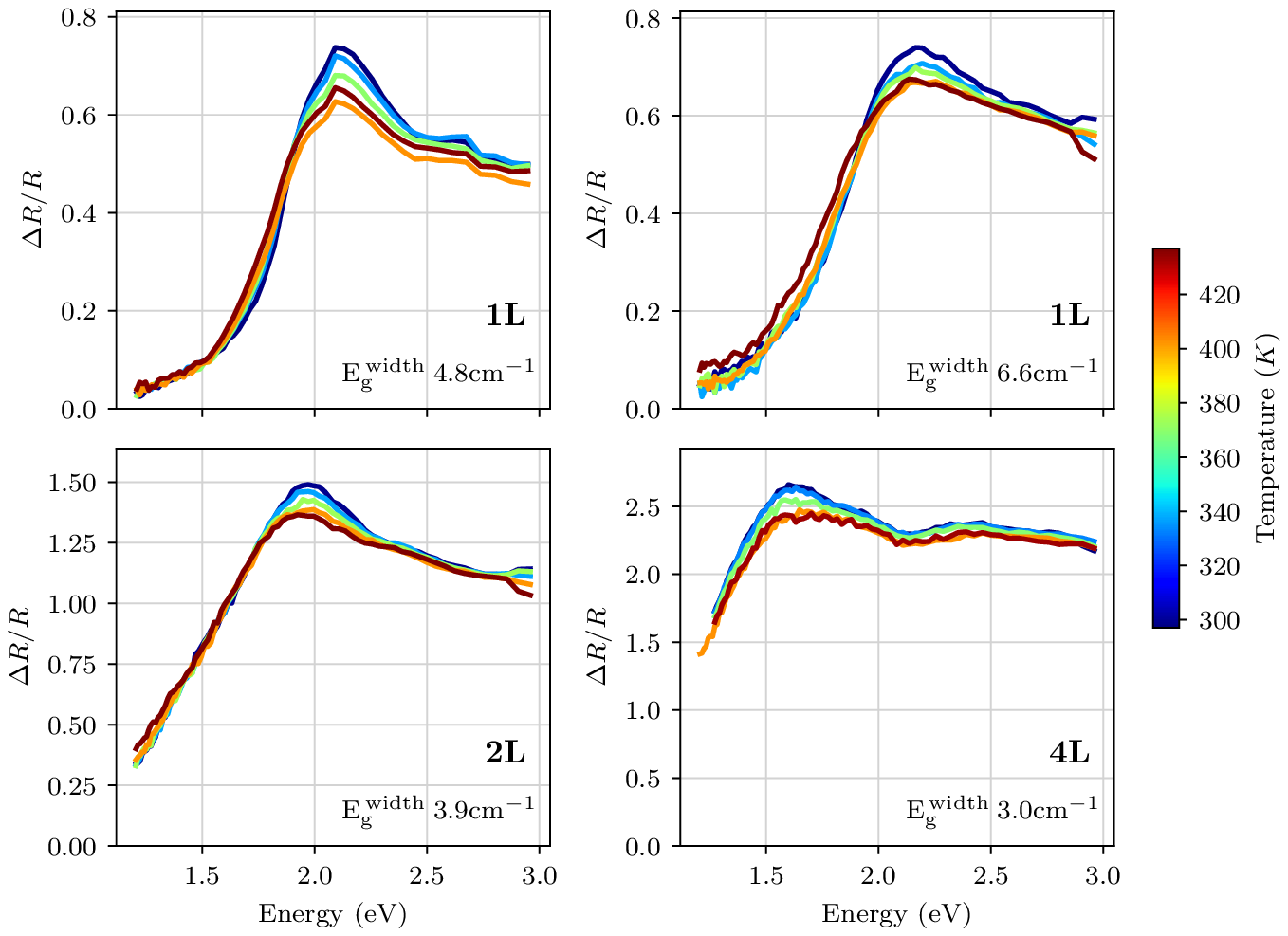}
\end{center}
\vspace{-5ex}
\caption{Differential reflectance of 1L, 2L and 4L flakes, featuring various samples quality characterized by the Raman $E_g$ mode linewidth \cite{articlePtSe2Raman}. Temperature is varied in the range $290 - 440\,\rm{K}$. Although the two monolayer samples feature different qualities, their temperature dependencies appear to be similar.}
\label{SM-fig:thermalDiffReflection}
\end{figure}

In order to vary the sample's temperature, we adapt the sample holder to include a planar resistive heater. The temperature can be varied between $290\,\rm{K}$ and $440\,\rm{K}$, and the presence of the heater allows only the reflection channel to be collected, not the transmission channel. Differential reflectance spectra $\Delta R/R = (R_{sample} - R_{substrate})\,/\,R_{substrate}$ are acquired and displayed in Fig. \ref{SM-fig:thermalDiffReflection}.

One can relate the differential reflectance with the real 2D optical conductivity, as detailed in \cite{LiHeinz2018OptConduct}, under the assumption of very low optical thickness $\left|Z_0 \sigma^{\mathrm{s}}\right| \ll 1$. This is what is used to compute the real optical conductivity for 1L and 2L $\rm{PtSe_2}$ in main text Fig. \ref{fig:absorptionTail}a.

\subsection{Monolayer \texorpdfstring{$1.0\,\rm{eV}$}{1.0 eV} peak} \label{SM-sec:1LPeak}

\begin{figure}[h!]
\begin{center}
\includegraphics[width=5in]{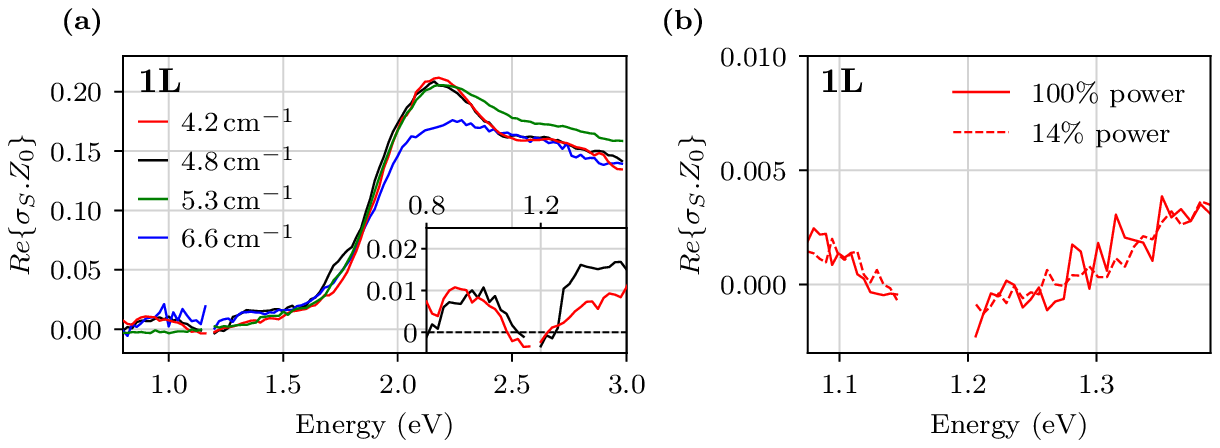}
\end{center}
\vspace{-5ex}
\caption{Monolayer $\rm{PtSe_2}$ low energy absorption peak. (a) 1L real 2D optical conductivity, labeled by the Raman $E_g$ mode linewidth \cite{articlePtSe2Raman}. (b) The same quantity on a more restricted range, for the highest-quality sample ($4.2\,\rm{cm^{-1}}$), with the laser operating at full power (full line), and at $14\,\%$ (dashed line).}
\label{SM-fig:1LVaryPower}
\end{figure}

Monolayer $\rm{PtSe_2}$ features a small peak at $1.0\,\rm{eV}$ (Fig. \ref{SM-fig:1LVaryPower}a) for the highest quality samples -- the ones featuring the lowest $E_g$ Raman mode linewidth \cite{articlePtSe2Raman}, of $4.2$ and $4.8\,\rm{cm^{-1}}$. One could suspect such peak to originate from the nature of the supercontinuum laser, which is based on a strong pulsed pump at $1.19\,\rm{eV}$ ($1064\,\rm{nm}$). However reducing the laser power to $14\,\%$ of its maximum setting does not modify the optical absorption close to this wavelength, ruling out the possibility for this $1.0\,\rm{eV}$ peak to be an artifact.

\section{Simulation of the optical absorption}
\label{SM-sec:theory}

\subsection{Density functional theory of the 1L to 12L band structure} \label{SM-sec:DFT}

Structural relaxation, electronic band structure and density of states computations are performed within the Density Functional Theory (DFT) formalism by using the Quantum Espresso (QE) package, which is an integrated suite of open-source codes for the electronic structure calculation and materials modeling at the nanoscale \cite{kohn1965self, giannozzi2009quantum,giannozzi2017advanced}. The result of DFT calculations depends on the choice of exchange-correlation functional (LDA, GGA, meta-GGA, etc.). This choice is done according to the specificities of the system under study, and more complex functionals are more computation-intensive, calling for a trade-off between accuracy and computational speed.
In our DFT calculation, the norm-conserving pseudo-potential with the Perdew-Burke-Ernzerhof (PBE) version of the generalized gradient approximation (GGA) is used to treat the exchange correlation interactions between valence electrons \cite{perdew1996generalized,van2018pseudodojo}. Since $\rm{PtSe_{2}}$ has a layered structure, the van der Waals (vdW) force between different layers are taken into account by first-principles calculations for a better accuracy \cite{grimme2004accurate,grimme2010consistent,thonhauser2015spin,tkatchenko2009accurate}.
In our work, a vdW correction of the PBE functional is therefore performed using Tkatchenko-Scheffler (TS) method \cite{tkatchenko2009accurate}.  The initial lattice constant for bulk $\rm{1T-PtSe_{2}}$ is taken from the materials project website \cite{jain2013commentary}, and is optimized for the chosen pseudo-potential, by using the quasi-Newton schemes as implemented in the QE package \cite{fischer1992general}. The optimized lattice constants of the 1T bulk are then used as initial parameters for the layered structures from 1L up to 12L and then relaxed.
As DFT relies on periodic boundary conditions, a vacuum space of at least $15\,\text{\AA}$ between periodic images of the structure along the [001] direction is used in order to avoid spurious couplings.
The cutoff energy and $\mathbf{k}$-point sampling were tested with a PBE-GGA calculation in a convergence study to ensure numerical stability. The kinetic energy cutoff for a plane-wave basis set was taken as $680\,\rm{eV}$. An appropriate Monkhorst-Pack $\mathbf{k}$-point sampling ~\cite{monkhorst1976special} of the Brillouin zone is used, centered with $15 \times 15 \times 1$, $20\times 20\times 1$, and $40\times 40\times 1$ meshes (for structure, self-consistent calculation, and projected density of states, respectively). The convergence criterion of self-consistent calculations for electronic structure and ionic relaxations is to reach a $10^{-8}\,\rm{eV}$  difference between two consecutive steps. During the optimization process, all atoms were free to move in all directions of space to minimize the internal forces down to $10^{-5}$ Rydbergs per Bohr radius.

\begin{figure}[h!]
\begin{center}
\includegraphics[width=7in]{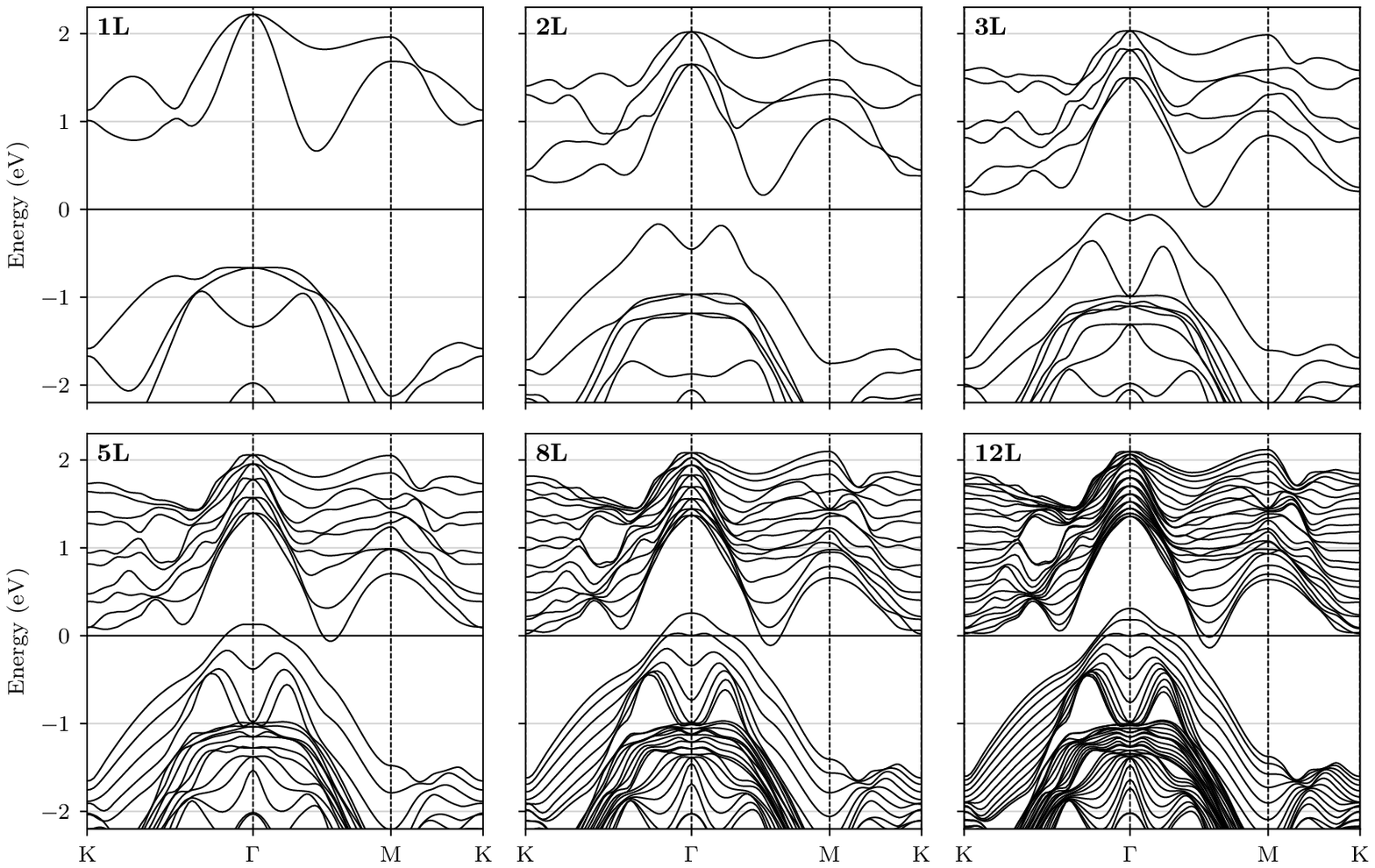}
\end{center}
\vspace{-6ex}
\caption{Band diagrams computed with GGA-PBE approximation for 1, 2, 3, 5, 8 and 12 layers $\rm{1T-PtSe_2}$.}
\label{SM-fig:bandsLayers}
\end{figure}

The 1T polytype of $\rm{PtSe_{2}}$ crystallizes in the centrosymmetric $CdI2$-type structure with space group $P\Bar{3}m1$  and point group $D_{3d}$ ($-3m$) \cite{lee1988development,guo1986electronic}. The structure can be regarded as hexagonal closely packed Se atoms with Pt atoms occupying the octahedral sites in alternate Se layers. The adjacent Se layers are held together by weak van der Waals interactions \cite{zhang2017experimental}.  The optimised lattice constant of the primitive unit cell of the monolayer $\rm{1T-PtSe_{2}}$  structure is found to be $3.72\,\text{\AA}$. Atomic positions in the monolayer form do not deviate appreciably from the bulk -- featuring a lattice constant of $3.78\,\text{\AA}$ -- indicative of strong bonding between the Pt and Se atoms. Lattice constants are detailed in table \ref{SM-tab:latticeConstants} and are in agreement with previous works \cite{Kandemir2018RamanDFT,li2016tuning,wang2015monolayer}.
The mean Pt - Pt spacing  is comparable with the single layer spacing measured by AFM on the exfoliated crystals (section \ref{SM-sec:AFM})

 \begin{table}[h!]
    \begin{tabular}{|C{6em}|C{3em}|C{3em}|C{3em}|C{3em}|C{3em}|C{3em}|}
    \hline
     & 1L  &  2L  & 3L & 5L & 8L & 12L \\
     \hline
     $a$ ($\text{\AA}$)  & 3.72 & 3.74 & 3.75  & 3.76 &  3.77 & 3.78 \\
    \hline
     $\langle d_{Pt-Pt}\rangle$ ($\text{\AA}$)  & - & 5.17  &  5.13 & 5.11 & 5.10  & 5.08 \\
      \hline
    \end{tabular}
    \label{SM-tab:latticeConstants}
    \vspace{2ex}
        \caption{Optimized lattice parameter $a$, and average Pt-Pt distance $d_{Pt-Pt}$ for each layer count.}
\end{table}

\subsection{Optical conductivity computation}

\subsubsection{Simulation and formula} \label{SM-sec:OptCondDerivation}

After obtaining the DFT-Kohn Sham (KS) eigensystem using QE, we use the Yambo code \cite{marini2009yambo, sangalli2019many, yamboCheatsheet2019}, to compute the direct interband dipolar matrix elements, involved in the linear optical response. They take the form $\left \langle m, \mathbf{k}|\hat{r}_{\alpha}| n, \mathbf{k}\right\rangle$, with $ \vert n, \mathbf{k}\rangle$ the single particle Bloch function of the $n$-th band obtained by DFT-KS computation for the wavevector $\mathbf{k}$, and $\hat{r}_{\alpha}$ the position operator along the $\alpha$ direction. These matrix elements contain all the symmetries imposed by the selection rules. For this computation, a plane-wave basis set is employed with a cutoff energy of $680\,\rm{eV}$, and a total of $50$ bands are included to ensure convergence of all computed quantities.

We consequently compute the diagonal imaginary dielectric permittivity using the dipolar matrix elements. Rather than using the bare output of the Yambo Fortran code, we prefer to use a simpler implementation that allows to extract contributions (see thereafter), and which faithfully reproduces the Yambo output for 1 to 12L, with a $< 3\%$ relative accuracy:
\begin{equation}
        Im\{\varepsilon_{\alpha, \alpha}\}    
    =  \frac{e}{\hbar}\frac{16}{\Omega} \sum_{n,m,\mathbf{k}} f_{n, \mathbf{k}}\left(1-f_{m, \mathbf{k}}\right) \times
     \left| \left\langle m, \mathbf{k}\left| \hat{r}_{\alpha}\right| n, \mathbf{k}\right\rangle\right|^2 
     \times Im\left\{\frac{2(E_{m,\mathbf{k}} - E_{n,\mathbf{k}})}{\left(E_{m,\mathbf{k}} - E_{n,\mathbf{k}}\right)^2 - \left(\omega + i\Gamma\right)^2}\right\}
\end{equation}
with $\Omega$ the unit cell volume, $d$ the cell out-of-plane extent, $f_{n\mathbf{k}}$ and $E_{n,\,\mathbf{k}}$ the occupation factor and energy of the $ \vert n, \mathbf{k}\rangle$ state, and $\Gamma$ the inhomogeneous broadening. The $e/\hbar$ factor converts the expression from Hartree atomic units to SI units. From the imaginary part of the dielectric function of this unit cell of vertical extent $d$, we compute the in-plane real 2D optical conductivity of a single cell, normalized by the impedance of free space $Z_0$, and averaged over the $x$, $y$ polarizations:
\begin{equation}
\begin{split}
    Re\{\sigma_{2D}.Z_0\} &= d\,Z_0\times Re\{\sigma\}\\
    &= d\,\frac{1}{\varepsilon_0 c}\times\frac{1}{2}\left( Re\{\sigma_{x,x}\}+Re\{\sigma_{y,y}\}\right)\\
    &= \frac{1}{2}\,\frac{d\,\omega}{c}\,\left( Im\{\varepsilon_{x,x}\}+Im\{\varepsilon_{y,y}\}\right)\\
\end{split}
\end{equation}
This finally leads to the expression:
\begin{multline}
    Re\{\sigma_{2D}.Z_0\} = \frac{16\,a_0\,e\,\omega}{\hbar\,c\,N_k} \sum_{n,m,\mathbf{k}} f_{n, \mathbf{k}}\left(1-f_{m, \mathbf{k}}\right)
   \times Im\left\{\frac{2(E_{m,\mathbf{k}} - E_{n,\mathbf{k}})}{\left(E_{m,\mathbf{k}} - E_{n,\mathbf{k}}\right)^2 - \left(\omega + i\Gamma\right)^2}\right\} \\
    \times\frac{1}{2}\left( \left| \left\langle m, \mathbf{k}\left| r_{x,x}\right| n, \mathbf{k}\right\rangle\right|^2 + \left| \left\langle m, \mathbf{k}\left| r_{y,y}\right| n, \mathbf{k}\right\rangle\right|^2\right)
\end{multline}
with $N_k$ the number of $\mathbf{k}$-points in the unit cell surface and $a_0$ the Bohr radius.

\begin{figure}[h!]
    \centering
    \includegraphics[width=5.2in]{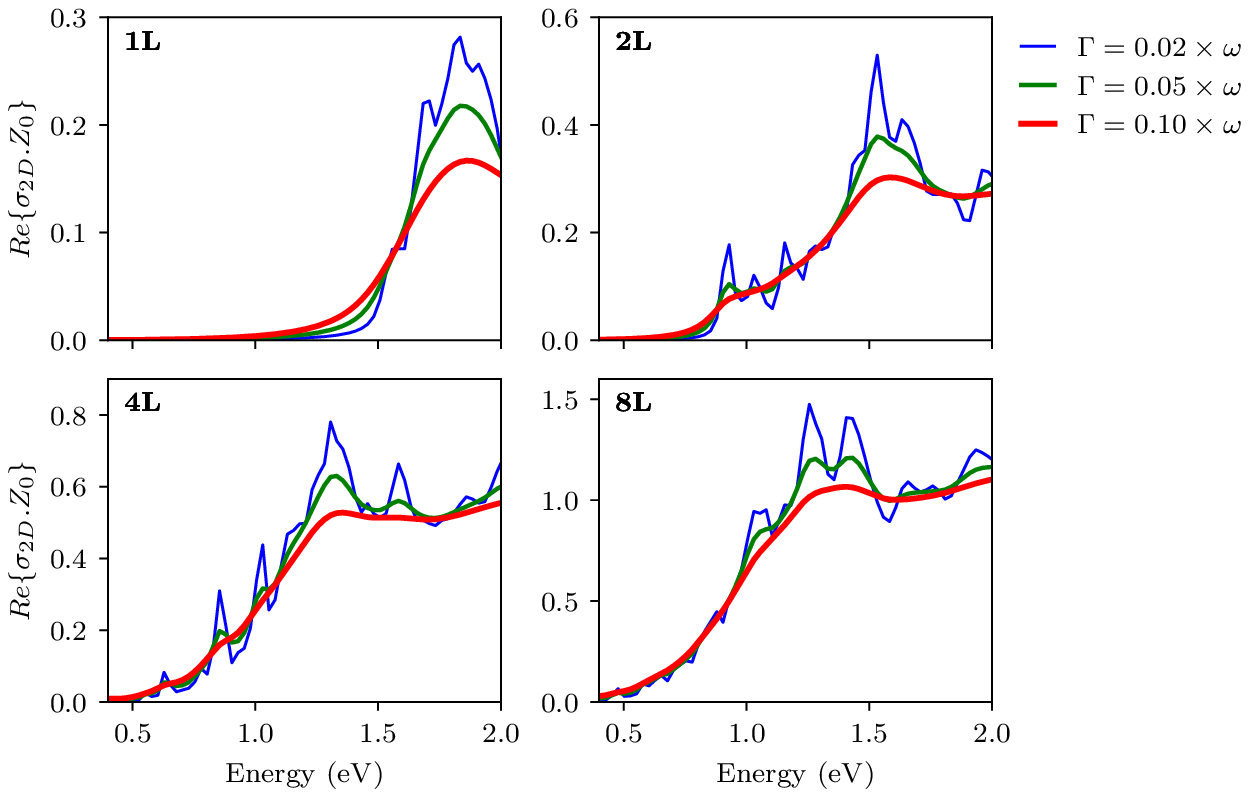}
\vspace{-2.5ex}
    \caption{Real 2D optical conductivity of 1L, 2L, 4L and 8L samples for increasing broadening $\Gamma = \{0.02,\,0.05,\,0.1 \}\times\omega$, with $\omega$ is the energy. In-tail resonant transitions appear while the broadening is reduced.}
    \label{SM-fig:absBroadening}
\end{figure}

From this expression one can identify the joint density of states, which is used in the main text Fig. \ref{fig:absorptionTail}c:
\begin{equation}
        jDOS(\omega) = \frac{16}{N_k} \sum_{n,m,\mathbf{k}} f_{n, \mathbf{k}}\left(1-f_{m, \mathbf{k}}\right)
   \times Im\left\{\frac{2(E_{m,\mathbf{k}} - E_{n,\mathbf{k}})}{\left(E_{m,\mathbf{k}} - E_{n,\mathbf{k}}\right)^2 - \left(\omega + i\Gamma\right)^2}\right\}
\end{equation}
In practice, the broadening $\Gamma$ is non-uniform, and we set it to increase linearly with energy  $\omega$, such that $\Gamma = 0.1\times\omega$. The effect of broadening on the optical absorption is illustrated in Fig. \ref{SM-fig:absBroadening}.
For the study of the many-body effects on monolayer optical conductivity in main text Fig. \ref{fig:gapExtrapolation}a, the broadening parameter $\Gamma$ is instead set to a fixed value of $0.2\,\rm{eV}$ for simplicity.

\subsubsection{Contributions diagram} \label{SM-sec:contributionDiagr}

\begin{figure}[h!]
    \centering
    \includegraphics[width=7in]{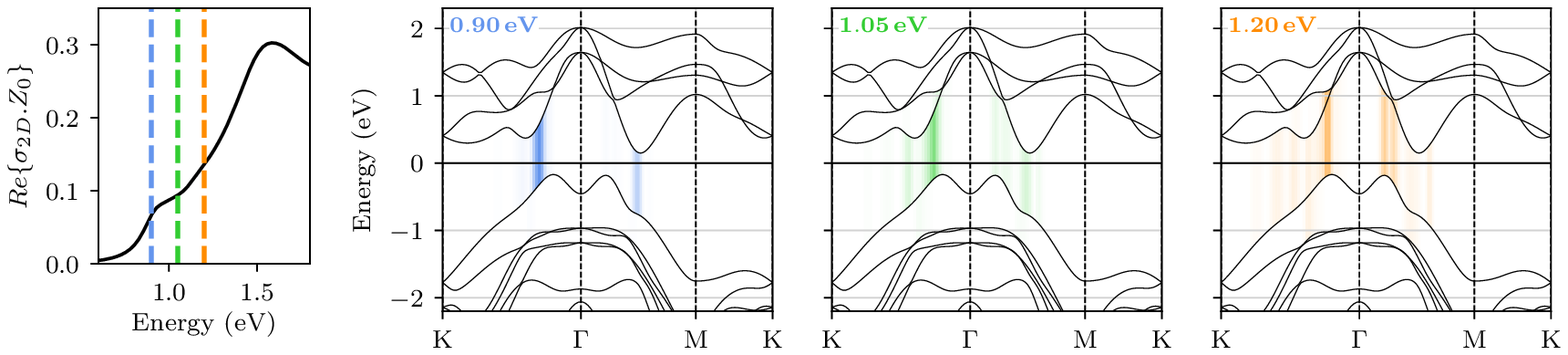}
\vspace{-6ex}
    \caption{Contribution diagram of the 2L $\rm{1T-PtSe_2}$ real 2D conductivity. At the shoulder energy of $0.9\,\rm{eV}$, Se $p_{x/y}^+$ and Pt $d_{zx/zy}$ orbitals come into play (SM-\ref{SM-sec:orbitals}).}
    \label{SM-fig:contributions2LTail}
\end{figure}

The computed real 2D optical conductivity can be inspected by displaying on the band diagram the relative contribution to the absorption of each pair of states -- each $n,m,\mathbf{k}$ (example in Fig. \ref{SM-fig:contributions2LTail}).

In practice, we use an interpolation of the k-mesh to estimate the contribution to the optical absorption of each k-point in the K, $\Gamma$, M and K path.
The relative weights of each point is taken into account to provide a faithful representation: a single $\Gamma$ is indeed shared by 12 copies of the same irreducible Brillouin zone (as pictured in main text Fig. \ref{fig:peaksContributions}a).
One can note that contribution diagrams display fringed patterns, as it is the case in Fig. \ref{SM-fig:contributions2LTail}. This may be due to an oscillating phase-matching of the states involved in the dipolar coupling.

\subsection{Further corrections applied to monolayer \texorpdfstring{$\rm{PtSe_2}$}{PtSe2}} \label{SM-sec:DFTCorrections1L}

Atomically thin two-dimensional (2D) materials host a rich set of electronic states that differ substantially from those of their bulk counterparts due to quantum confinement and enhanced many-body effects \cite{thygesen2017calculating}. It is known that a quantitative description of the optical response of an interacting electron system should account for electron–hole interactions. The description of these interactions goes far beyond the standard DFT scheme and the free quasiparticle picture, since electron-hole pairs can bind in bound excitons.  One common approach to  compute quasiparticle band structure and optical response while including electron-electron interactions and excitonic contributions is the GW plus Bethe Salpeter equation (GW-BSE), within the many body perturbation framework \cite{sangalli2019many,RevModPhys.74.601,PhysRevB.34.5390,strinati1988application}. This method has been applied with success to a wide variety of materials, including systems with reduced dimensionality such as $\rm{MoS_2}$ \cite{qiu2013optical}, carbon nanotubes \cite{spataru2004excitonic} and graphene \cite{yang2009excitonic}.

\subsubsection{From GGA functional to GW}
\label{sec:GW}

Due to the well-known tendency of the semi-local exchange-correlation functional used in standard DFT (the local density (LDA) or generalized gradient (GGA) approximations, etc.) to underestimate bandgaps, we use many-body methods, by means of the quasiparticle (QP) concept (GW approximation), to correct the bandgap of the DFT band structure. Physically, the QP energies describe the energy cost of adding electrons/holes to the neutral ground state of the material. In fact, this approximation incorporates screened electron-electron correlation, which in particular permits substantial amelioration of the energy bandgaps with respect to experiments \cite{ sangalli2019many,rangel2020reproducibility,van2006quasiparticle,guandalini2023efficient}.

In our work, the GW calculations is performed using the many body perturbation theory (MBPT) open-source code package Yambo \cite{onida2002electronic,strinati1988application,marini2009yambo, sangalli2019many}. We employ the "one-shot" $\rm{G_0 W_0}$ approximation. The inverse of the microscopic dynamic dielectric function is obtained within the plasmon-pole approximation \cite{sangalli2019many,larson2013role}.

We perform quasi-particle band structure simulations on a $30 \times 30 \times 1$ k-points grid using up to $200$ empty bands, and we truncate exchange and correlation energies at $60\,\rm{Ry}$ and $13\,\rm{Ry}$, respectively. 

\begin{figure}[h!]
\begin{center}
\includegraphics{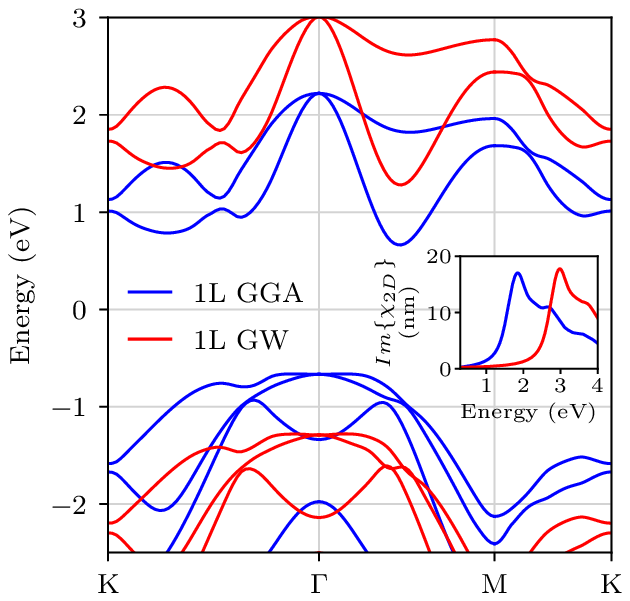}
\end{center}
\vspace{-6.5ex}
\caption{Band diagrams computed with GGA-PBE and GW methods for monolayer $\rm{1T-PtSe_2}$. The many-body corrections result in a rigid shift of the conduction bands with respect with the valence bands, with very little modification of the bands' profiles. In inset is plotted the 2D imaginary surface susceptibility $Im\{\chi_{2D}\} = Re\{\sigma_{2D}\}\,/\,\varepsilon_0\omega = d\,(Im\{\varepsilon\}-1)$ \cite{LiHeinz2018OptConduct}. The bands rigid shift appears as an energy shift in susceptibility.}
\label{SM-fig:Bands1LGGAGW}
\end{figure}

To speed up convergence with respect to empty states, we adopt the technique described in Ref \cite{bruneval2008accurate} as implemented in the YAMBO code. We have numerically verified the convergence of the QP gap up to $50\,\rm{meV}$ with respect to the number of bands included in the calculation of both the Green and polarization functions. The Coulomb interaction is truncated in the layer-normal direction to avoid spurious interactions with the image systems. GGA and GW band diagrams are represented in Fig. \ref{SM-fig:Bands1LGGAGW}.

\vspace{-2ex}

\subsubsection{Bethe-Salpeter Equation correction}

The Bethe-Salpeter Equation (BSE) is solved  using the Tamm-Dancoff approximation \cite{dancoff1950non,onida2002electronic} and takes into account the local field effects as implemented in yambo code \cite{marini2009yambo, sangalli2019many,guandalini2023efficient}. The BSE is calculated on top of the GW eigenvalues. We obtained converged excitation energies considering, eight empty states and eight occupied states in the excitonic Hamiltonian, the irreducible BZ being sampled up to a $30\times 30 \times 1$ k-points mesh. The same cut-offs of GW are used to build up the exchange electron-hole attractive and repulsive kernels in the BSE matrix. It is important to note that electronic transitions are very sensitive to BZ samplings. Altogether, such a treatment is much more intricate as compared to standard DFT scheme, and often restricted to systems that are small in size, due to its exceptional computational cost. For these simulation, a constant broadening of the interband transitions is used (see \ref{SM-sec:OptCondDerivation}), of $0.2\,\rm{eV}$.

\subsubsection{Hybrid functional HSE06}
\label{SM-sec:HSE-RigidShift}

The DFT-PBE approximation often underestimates the fundamental energy band gaps due to self-interaction errors. This issue arises from the discontinuity of the chemical potential in the exchange-correlation potential, which leads to inaccuracies in the prediction of electronic band gaps \cite{heyd2005energy, perdew1996comparison,sham1983density}. A manner to overcome this shortfall is to use Hybrid functional, which is extensively used to investigate a variety of periodic systems with plane wave basis sets. We shall see that the hybrid functional leads to different estimation of the band gap at the cost of a large increase in computation time. Fortunately, the relevant portion of the band structure as well as optical absorption are strongly similar to GGA. 

Among the several proposed  hybrid functional (PBE0 \cite{perdew1996rationale}, B3LYP \cite{becke1988density}, HSE \cite{heyd2003hybrid} etc.), in our work we will use the Heyd-Scuseria-Ernzerhof functional (HSE06), which is  often used to compute the electronic properties of layered materials. HSE06  is a type of exchange-correlation functional which combines the KS orbitals (GGA/LDA) with a portion of exact Hartree-Fock (HF) exchange.  Specifically, HSE06 splits the exchange interaction into short-range and long-range components, applying the HF exchange only to the short-range part and using a GGA-like exchange for the long-range part \cite{becke1993new,heyd2003hybrid}.

The electronic structure calculations were carried out using the hybrid functional HSE06 implemented in QE \cite{kohn1965self, giannozzi2009quantum,giannozzi2017advanced}. The HSE06 functional incorporates $25\%$ exact HF exchange.  For the plane-wave basis set, a cutoff energy of $60 R_{y}$ for the wavefunctions and a k-point grid of $15 \times 15\times 1$  for Brillouin zone was used.  The q-grid dimensions for the HF exchange interaction were set to $10\times 10 \times 1$. These parameters are crucial for ensuring the convergence of the HF exchange interaction in HSE06 calculations, which directly impacts the accuracy of the hybrid functional results. However, increasing these q-grid dimensions can significantly raise the computational cost, as it requires more extensive calculations to accurately capture the exchange interactions. Wannier90 
\cite{marzari2012maximally, pizzi2020wannier90} was then used to  project the electronic states onto localized Wannier functions and to map band structure energies from k-grid-based calculations onto finely spaced k-paths by means of Wannier interpolation.  

\begin{figure}[h!]
\begin{center}
\includegraphics{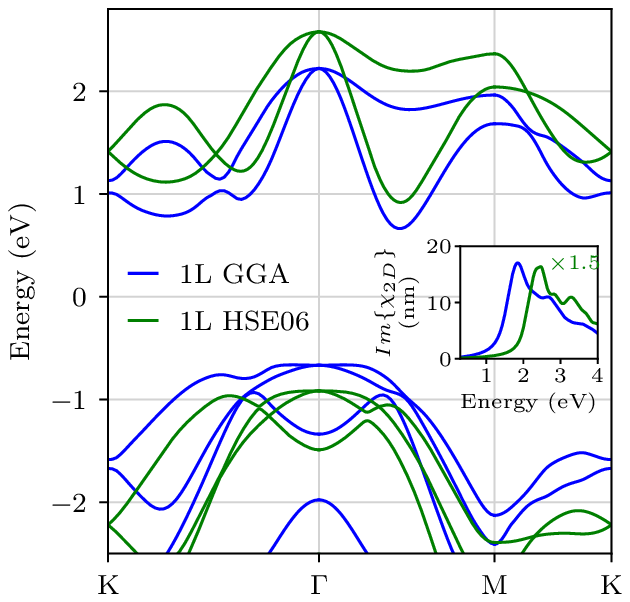}
\end{center}
\vspace{-6.5ex}
\caption{Band diagrams computed with GGA-PBE and HSE06 methods for monolayer $\rm{1T-PtSe_2}$. In inset is plotted the imaginary surface susceptibility $Im\{\chi_{2D}\} = Re\{\sigma_{2D}\}\,/\,\varepsilon_0\omega = d\,(Im\{\varepsilon\}-1)$ \cite{LiHeinz2018OptConduct}.}
\label{SM-fig:Bands1LGGAHSE}
\end{figure}

The electronic band structure using GGA and HSE functional for 1ML PtSe2 are presented in Figure \ref{SM-fig:Bands1LGGAHSE}. The indirect band gap values determined by the Wannier-interpolated HSE06 band structure is increased by 0.5 eV as compared to GGA approximation, besides, the band structure is unchanged in the band edges of CBM and VBM. 
Computationally, using HSE06 is more demanding than GGA due to the need to compute the exact exchange integrals, which require a non-local treatment of the electron-electron interaction. This increases the computational cost and complexity, often necessitating more sophisticated algorithms and longer computation times, particularly for multilayer cases. Therefore, in the main text we choose to use the GGA approximation.

\subsubsection{Finite temperature optical absorption}
\label{SM-sec:DFT-temperature}

To explore the role of phonons in the optical absorption spectrum and the band-gap renormalization, we employed the method of special displacements implemented in the  \textit{ZG.x} code within EPW \cite{ponce2016epw}. This nonperturbative approach involves computing the band structure and optical spectra at each temperature using a sufficiently large supercell. The atomic positions in this supercell are appropriately distorted according to the phonon displacement patterns calculated using density-functional perturbation theory (DFPT). Below, we summarize the calculation procedure and convergence parameters. A detailed description can be found in the following references \cite{ponce2016epw,Zacharias2020SpecialDisplacement,Zacharias2016OneShotPRB}

(i) Using DFPT module of Quantum Espresso,  We determine the interatomic force constants  in the primitive unit cell, using a uniform grid of  $8 \times 8 \times 1$ q-points. By diagonalizing the dynamical matrix obtained from the matrix of force constants, we determine the phonon eigenmodes $e_{\kappa \alpha,\nu} (\boldsymbol{q})$ and eigenfrequencies $\omega_{\boldsymbol{q} \nu}$ ($\kappa$ and $\alpha$ indicate the atom and the Cartesian direction, respectively) used  to construct the Zacharias Guistino (ZG) displacement. 
The reliable determination of ZG displacement requires the knowledge phonon-band dispersion with an accuracy on the order of $0.1\ \mathrm{meV}$. Therefore, we perform the DFPT calculation with a high cutoff energy of $100R_{y}$.

(ii) In preparation for calculations on a supercell of size $8\times8 \times 1$  , we interpolate the vibrational eigenmodes and frequencies on a finer q-points grid with the same size as the supercell, then at each temperature, a $8\times8 \times 1$ supercell of 179 atoms is constructed starting from the real-space force constants, by using the  \textit{ZG code}  that is part of EPW open source package \cite{ponce2016epw}. 

(iii) To compute band structures, we first set up the k-point path in the Brillouin zone of the primitive cell, then we obtain the folded k-points in the Brillouin zone of the supercell. We perform a DFT band structure calculation in the supercell, and we unfold the result using the method of Ref. \cite{popescu2012extracting}. Supercell GGA  calculations are performed with wavefunction cutoff of $80 R_{y}$.

(iv) To obtain temperature-dependent  optical absorption spectra, steps (i)– (ii) above remain the same, while step (iii) is  replaced by the calculation of the optical matrix elements in the independent particle approximation (IPA).


\begin{figure}[h!]
\begin{center}
\includegraphics{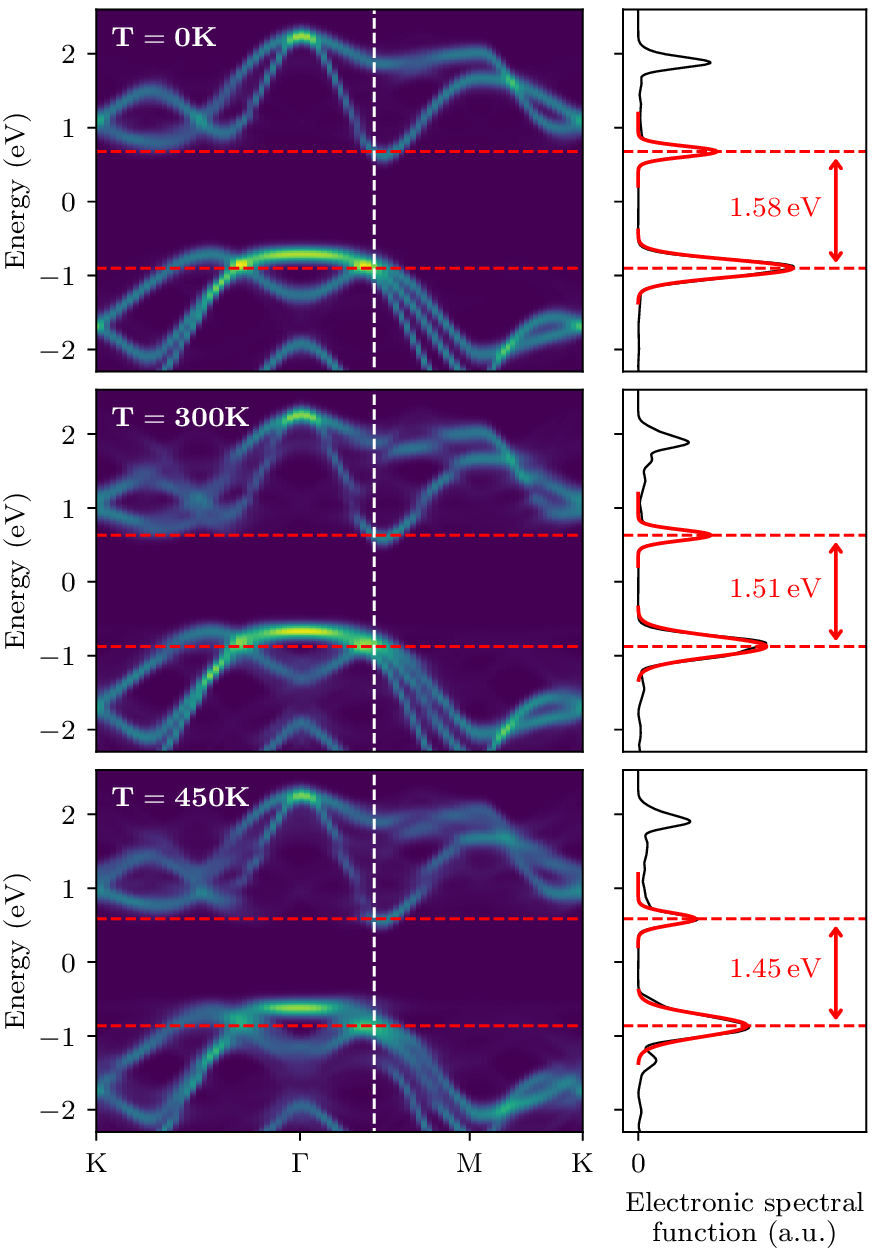}
\end{center}
\vspace{-6.5ex}
\caption{(left) Spectral function of ML PtSe$_{2}$ calculated using the special displacement method as implemented in the ZG code at $T = 0\,\mathrm{K}$, $300 \,\mathrm{K}$ and $450 \,\mathrm{K}$. (right) Spectral function at the direct bandgap \textbf{k} wavevector (dashed white line in the left panel). The direct bandgap value is extracted in between the conduction and valence bands using Gaussian fitting of the spectral function.}
\label{SM-fig:bandDiagShiftTemp}
\end{figure}

To assess the role of temperature renormalization of the electronic band structure, which may induce a shift of the bands and can be mistaken for an effect of the indirect transitions, we plot in  Fig. \ref{SM-fig:bandDiagShiftTemp} the spectral function of $\mathrm{PtSe_{2}}$ for three different temperatures $T = 0\,\mathrm{K}$, $300 \,\mathrm{K}$ and $450 \,\mathrm{K}$ obtained from the ZG displacement \cite{Zacharias2016OneShotPRB} in a $8 \times 8 \times 1$ supercell.

By extracting the corresponding direct temperature-dependent bandgap, we find that it decreases from $1.58\,\mathrm{eV}$ at $T = 0\,\mathrm{K}$ to $1.45\,\mathrm{eV}$ at $T = 450\,\mathrm{K}$. This confirms a renormalization effect (redshift) with temperature, that explains the small redshift of the optical absorption tail observed in the temperature-dependent simulations (main text Fig. \ref{fig:absorptionTail}a).
This allows us to definitively exclude phonon-assisted indirect transitions as the main cause of the observed experimental absorption tail. 

\subsection{AB stacked bilayers } \label{SM-sec:ABStacking}

Stacking of the sublayers in a layered material directly affect its electronic and optical properties.
Different stackings of 2L $\rm{PtSe_2}$ can theoretically be obtained through layer shifting and/or rotation at various angles. In our case, we will study three possible stackings, namely $\rm{AA}$, $\rm{AB_1}$ and $\rm{AB_2}$.
 The $\rm{AA}$ stacking is the most stable form of 2L $\rm{PtSe_2}$ (used in main text Fig. \ref{fig:conductivitiesSchemes}), it corresponds to the stacking order of bulk $\rm{PtSe_2}$. In this structure, the Pt atoms are aligned vertically. The $\rm{AB_1}$ structure’s  is obtained by rotating the upper layer of $\rm{AA}$ $\rm{PtSe_2}$ by $180^{\circ}$ around the Pt atom. The $\rm{AB_2}$ structure is obtained by translating the upper layer of the $\rm{AB_1}$ structure in the direction of $\mathbf{a}-\mathbf{b}$ with a magnitude of $ \vert \mathbf{a}-\mathbf{b}\vert/3$.  After constructing $\rm{AB_1}$ and $\rm{AB_2}$ stacking both are fully relaxed on both the lateral cell constants and ion positions based on DFT as implemented in QE (see section  \ref{SM-sec:DFT}).
These bilayer $\rm{PtSe_2}$ calculations are based on the work of reference \cite{Fang2019AA_AB_DFT}.
\begin{table}[h!]

    \begin{tabular}{|L{6em}|C{4em}|C{4em}|C{4em}|}
    \hline & $\rm{AA}$  &  $\rm{AB_{1}}$  & $\rm{AB_{2}}$  \\
    \hline $a$ (\r{A})  & 3.74 & 3.69 &  3.71 \\
    \hline $d_{pt-pt}$ (\r{A}) &5.17  &6.42 &  5.38 \\
    \hline $\Delta E_{B}$ (meV) & -322 &-286 &  -282\\
    \hline
    \end{tabular}
    \label{tab:parameter}
        \caption{Optimized lattice parameter $a$, Pt-Pt distance $d_{Pt-Pt}$, and binding energy $\Delta E_{B}$ for different stackings 2L $\rm{PtSe_2}$.}
\end{table}

\begin{figure}[h!]
\begin{center}
\includegraphics[width=7in]{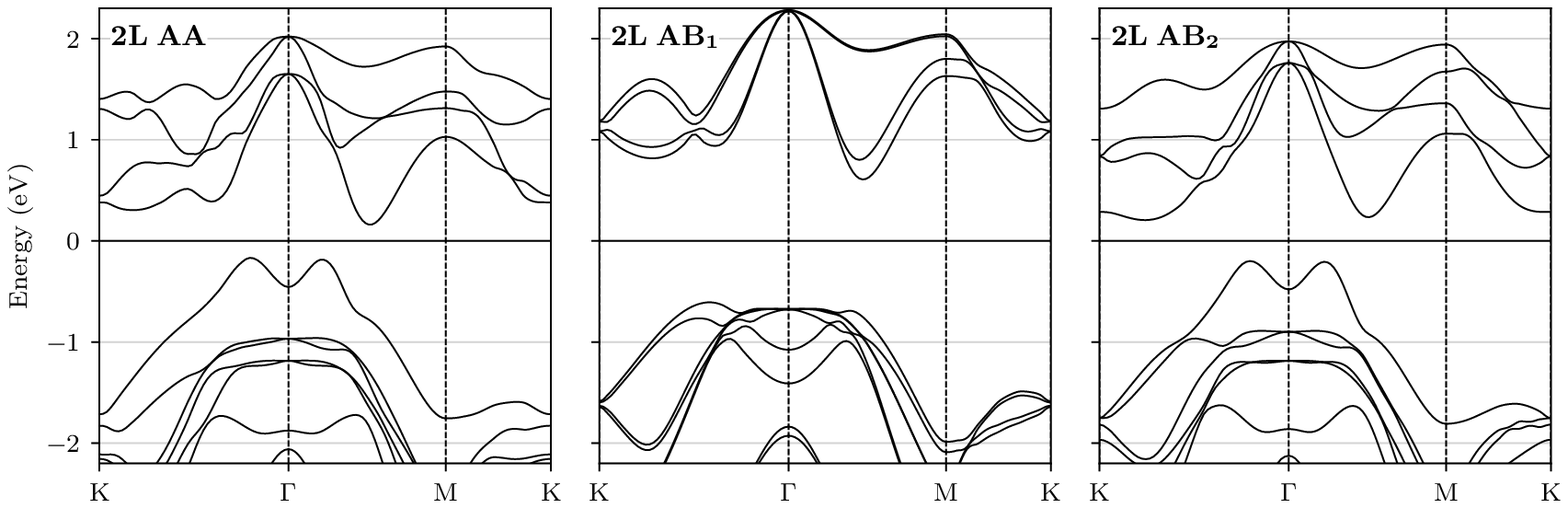}
\end{center}
\vspace{-6.5ex}
\caption{Band diagrams computed with the GGA method for 2L with $\rm{AA}$, $\rm{AB_1}$ and $\rm{AB_2}$ stacked $\rm{1T-PtSe_2}$.}
\label{SM-fig:bands2L}
\end{figure}

 In table \ref{tab:parameter}, we display the lattice constant, the interlayer distance $d_{Pt-Pt}$ and the Binding energy $\Delta E_{B}$ for the three mentioned structure. As expected, we find that the top-to-top ($\rm{AA}$) arrangement is  energetically the most stable stacking of 1T-$\rm{PtSe_2}$ layers. In Fig. \ref{SM-fig:bands2L},  we plot their band structure.

Different stackings could effectively modulate the band structures of bilayer $\rm{PtSe_2}$. The $\rm{AA}$ structure has the minimum bandgap of $0.3\,\rm{eV}$ and the larger interlayer distance, while the corresponding $AB_{1}$ -- which has the lower interlayer distance -- has a bandgap almost two times larger.

\subsection{Orbital projected partial density of states} \label{SM-sec:orbitals}
\subsubsection{Simulation method}

In order to identify the contribution of the different atomic orbitals to the electronic band structure and hence to the optical absorption spectra, we calculate the projected density of states (PDOS) by projecting the wave-functions onto the localized atomic orbitals.  Such a calculation is done using the Projector Augmented Wave (PAW) method, as implemented in Quantum Espresso's through projwfc.x code. A PAW pseudopotential, denser k point $40 \times 40 \times 1$ and the tetrahedron method with Blöchl's  correction was used \cite{kohn1965self, giannozzi2009quantum,giannozzi2017advanced}.

\subsubsection{Orbitals projections}
We investigate the electronic state projection onto the localized atomic orbitals. To do so, we monitor the valence states of the Selenium atoms 4s and 4p, and those of the Platinum atoms 5d, 6s and 6p. The orbitals components displayed thereafter are the projection coefficients modulus squared obtained from the PDOS method.

\begin{figure}[h!]
\begin{center}
\includegraphics[width=6in]{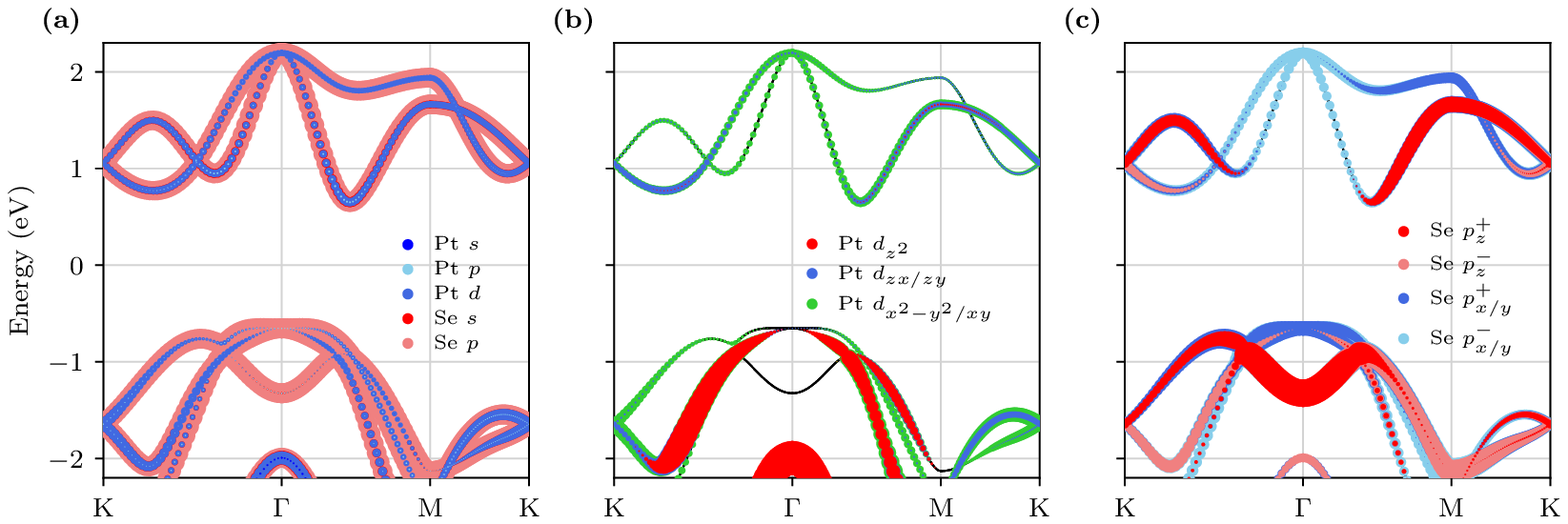}
\end{center}
\vspace{-5ex}
\caption{Projection of the electronic wavefunction of 1L $\rm{PtSe_2}$ on localized atomic orbitals of (a) Platinum and Selenium atoms, (b) $d$ orbitals of the Platinum atom, (c) the $p$ orbitals of the Selenium atoms, in symmetric and antisymmetric combinations.}
  \label{SM-fig:orbitalDecomp1L}
\end{figure}

The Se $p$ and the Pt $d$ orbitals are dominating the band diagram around the bandgap energy region, as displayed in Fig. \ref{SM-fig:orbitalDecomp1L}a, for the case of a monolayer. The Pt $d$ orbital can be decomposed further in $d_{z^2}$, $d_{zx/zy}$, $d_{x^2-y^2/xy}$ harmonics (Fig. \ref{SM-fig:orbitalDecomp1L}b), for which we gather the atomic orbitals in three groups accounting for Z-rotation symmetry, due to the unpolarized light in our problem. The Se $p$ orbital can likewise be decomposed in $p_{z}$ and $p_{x/y}$ harmonics. However there are two Selenium atoms per unit cell, and one can consider symmetric and antisymmetric combinations of the two states, thereafter labeled $+/-$ (Fig. \ref{SM-fig:orbitalDecomp1L}c).

\subsubsection{Orbitals decomposition of optical transitions}

We want to assess the coupling strength between these orbitals to understand the features of the optical absorption computed in our study.
To estimate the coupling of orbitals, we use a procedure similar to the Slater Koster calculation of interatomic elements for tight-binding models \cite{Slater1954tightBinding}.
This coupling is evaluated by the square of the dipolar matrix elements, as in the equation:

\begin{equation}
   \left|\langle \Psi_1 | \mathbf{p} | \Psi_2 \rangle \right| ^2 = \left|-i\hbar.\int \rm{d}\mathbf{r}\, \Psi_1(\mathbf{r})(\mathbf{\nabla}\Psi_2(r))\right|^2
\end{equation}

\begin{figure}[h!]
    \begin{center}
            \includegraphics[width=4in]{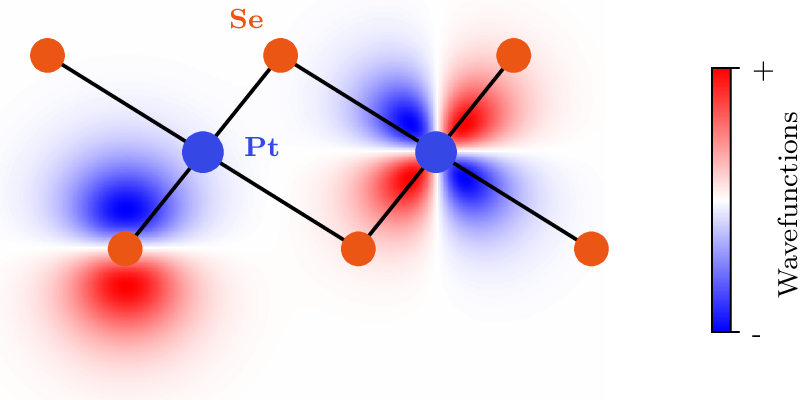}

    \end{center}
\vspace{-8ex}
    \caption{Se $p_z$ (left) and Pt $d_{zy}$ (right) orbitals anzatz, represented in the $yz$ plane, at the atomic positions, with $d = 0.5\,a_0$.}
    \label{SM-fig:orbitalsExemple}
\end{figure}

To be non-zero, the electromagnetic interaction has to involve orbitals of different parities (excluding Se $p$ - Se $p$ interaction for instance), Pt $d$ and Se $p$ coupling having therefore the prominent contribution.
We evaluate numerically the coupling of each Platinum $d$ orbital harmonics with the $p$ orbital harmonics of the 6 neighboring Selenium atom by using for the atomic states the following anzatz: 

\begin{equation}
    \Psi_1(\mathbf{r}) \sim r^l\,e^{-r/d}\,Y_{l,m}(\theta, \phi)
\end{equation}

where $l$ and $m$ are the azimuthal and magnetic quantum numbers, $Y_{l,m}$ the associated spherical harmonic and $d = 0.5\,a_0$ is the controlling the radial range ($a_0$ is the Bohr radius), chosen such that the orbitals spread is typically on the order of the distance between the Se and Pt planes (illustrated in Fig. \ref{SM-fig:orbitalsExemple}).

\begin{table}[h!]
    \centering
    \begin{tabular}{|l|C{4em}|C{4em}C{4em}|C{4em}|C{4em}C{4em}|}
    \hline
    & \multicolumn{3}{c|}{$x$ polarization} & \multicolumn{3}{c|}{$y$ polarization} \\
    \hline
    Orbitals & Se $p_{z^+}$ & Se $p_{x^+}$ & Se $p_{y^+}$ & Se $p_{z^+}$ & Se $p_{x^+}$ & Se $p_{y^+}$ \\
    \hline
    Pt $d_{z^2}\;$ & $0$ & $0.15$ & $0$ & $0$ & $0$ & $0.15$  \\
    \hline
    Pt $d_{zx}\;$ & $0.21$ & $0$ & $1.00$ & $0$ & $1.00$ & $0$  \\
    Pt $d_{zy}\;$ & $0$ & $1.00$ & $0$ & $0.21$ & $0$ & $1.00$  \\
    \hline
    Pt $d_{x^2-y^2}\;$ & $0$ & $0.01$ & $0$ & $1.00$ & $0$ & $0.01$  \\
    Pt $d_{yx}\;$ & $1.00$ & $0$ & $0.01$ & $0$ & $0.01$ & $0$  \\
    \hline
    \end{tabular}
    \vspace{2ex}
    \caption{Coupling Selenium and Platinum orbitals in arbitrary units. $x$ is being chosen along the main crystallographic axis. $x$ and $y$ polarizations couple complementary orthogonal orbitals with the same intensity, in agreement with the definitions of Pt $d_{z^2}$, $d_{zx/zy}$, $d_{x^2-y^2/xy}$ and Se $p_{z^+}$, $p_{x/y}^+$ orbitals groups to account for unpolarized light. Asymmetric combinations of Se $p$ orbitals where found to present a zero coupling with Pt $d$ orbitals.}
    \label{SM-tab:orbitalsCouplings}
\end{table}

\vspace{-2ex}
\begin{figure}[h!]
    \begin{center}
    \includegraphics[width=5in]{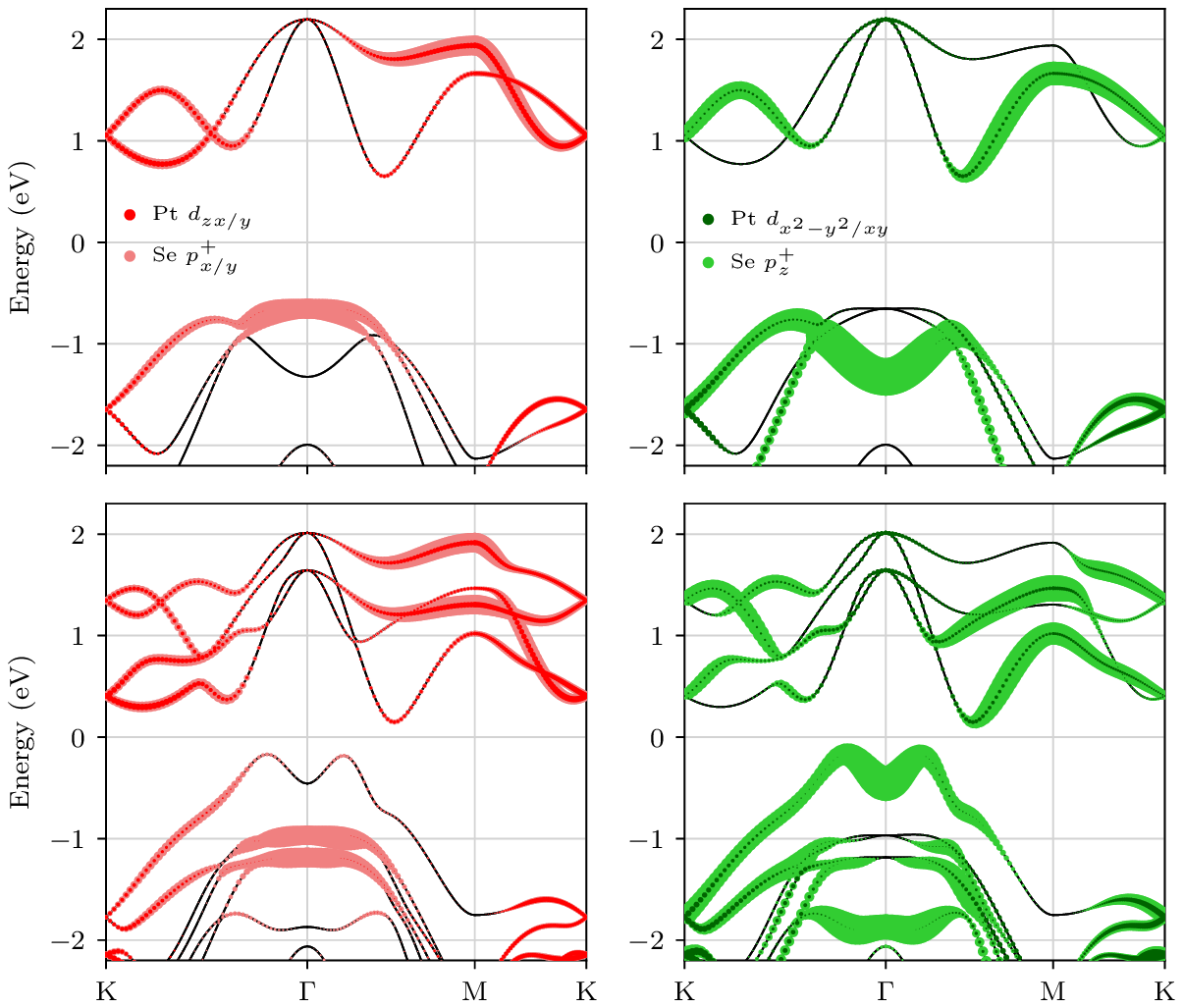}
    \end{center}
\vspace{-6.5ex}
\caption{Orbital decomposition of the main orbitals which coupling is responsible for the optical absorption: (left) Pt $d_{zx/zy}$ and Se $p_{x/y}^+$, and (right) Pt $d_{x^2-y^2/xy}$ and $p_{z^+}$, for (top) monolayer and (bottom) bilayer $\rm{PtSe_2}$.}
    \label{SM-fig:orbitalCouplings1L2L}
\end{figure}

We find that the couplings involving an antisymmetric combination of Selenium P orbitals are zero, which relies on crystal symmetries. The couplings involving its symmetric part are detailed in \textbf{table \ref{SM-tab:orbitalsCouplings}}, for either $x$ or $y$ light polarization. The computed values are indicative, and vary with the anzatz range of the atomic wavefunction. Nonetheless two dominating couplings are systematically standing out, with equal intensities: Pt $d_{zx/zy}$ and Se $p_{x/y}^+$ on one hand, and Pt $d_{x^2-y^2/xy}$ and $p_{z^+}$ on the other hand.
This justifies that we only consider the couplings of these orbitals as the principal source of optical absorption. The wavefunctions projections on these orbitals are consequently drawn in Fig. \ref{SM-fig:orbitalCouplings1L2L}, for the cases of 1L and 2L.

\section{Density of states, peaks and low-energy scaling}

\subsection{Optical absorption and jDOS} \label{SM-sec:absorptionPeaksJDOS}

\begin{figure}[h!]
    \centering
    \includegraphics[width=5in]{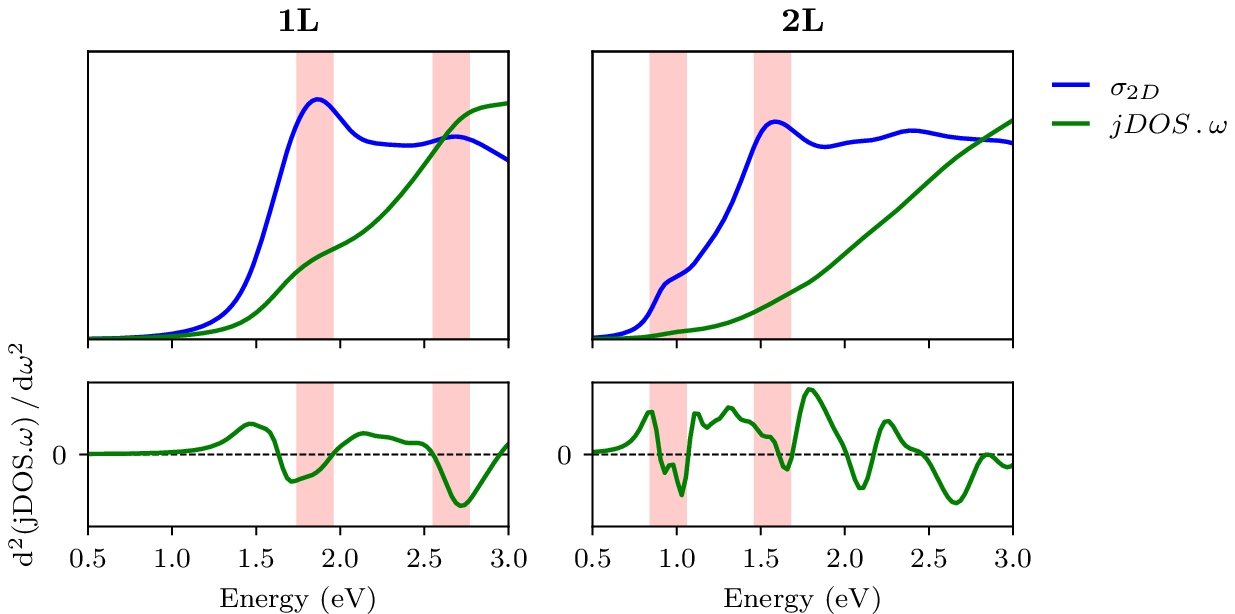}
\vspace{-2.5ex}
    \caption{Absorption peaks and jDOS. (top) Simulated real 2D optical conductivity and joint density of states $jDOS$, and (bottom) $jDOS$ curvature, for (left) monolayer and (right) bilayer stackings. In red are highlighted the main peaks energies: $1.85\,\rm{eV}$ and $2.66\,\rm{eV}$ for 1L, $0.95\,\rm{eV}$ and $1.57\,\rm{eV}$ for 2L.
}
    \label{SM-fig:conductivityPeaksJDOS}
\end{figure}

We come back to the nature of the absorption peaks, by inspecting the joint density of states. We already know that the dipolar couplings responsible for the absorption vary smoothly around the region of the Brillouin zone involved in the peak's optical absorption (Fig. \ref{SM-fig:orbitalCouplings1L2L}), and are thus unlikely to be the cause of such maxima.
To confirm that the optical absorption peaks are indeed due to the band nesting effect, we check that these peaks are present in the joint density of states. We compare the real 2D optical conductivity and the $jDOS$ times frequency ($\sigma_{2D} \sim\omega\,.\,M\,.\,jDOS$ where $M$ is the dipolar coupling, see section \ref{SM-sec:OptCondDerivation}), for 1L and 2L $\rm{PtSe_2}$ (Fig. \ref{SM-fig:conductivityPeaksJDOS} top).

If some shoulders are visible in the 1L $jDOS$, corresponding to the main peaks, they are hardly noticeable for the 2L. This is because the multiplicity of the bands average out the contributions to form a smoothly increasing function. We prefer to take the second derivative of $\omega\,.\,jDOS$ in order to distinguish the most-pronounced local maxima (they appear as minima of the double derivative, see Fig. \ref{SM-fig:conductivityPeaksJDOS} bottom). We can distinguish the two main peaks for 1L, and the main peak for the 2L as well as its small peak located in the absorption tail. However it is hard to conclude for the 2L second absorption peak, since several transitions come into play.

\subsection{Low-energy scaling law} \label{SM-sec:scalingLaw}

To understand how to extract a scaling from the energy bands' landscape to estimate the optical absorption tail, let's first recall that the 2D optical conductivity $\sigma_{2D}$ and the joint density of states $jDOS$ can by obtained using sums over the band diagram:

\begin{equation}
\sigma_{2D}(\omega) \sim \sum_{c,v}\int_\mathbf{k}\mathrm{d}^3\mathbf{k}\, M_{k,c,v}\: \delta[\omega - (E_{c,\mathbf{k}} - E_{v,\mathbf{k}})]
\end{equation}
\begin{equation}
jDOS(\omega) \sim  \sum_{c,v}\int_\mathbf{k}\mathrm{d}^3\mathbf{k}\,\delta[\omega - (E_{c,\mathbf{k}} - E_{v,\mathbf{k}})]
\end{equation}

where $\omega$ is the frequency, $c$ and $v$ designate the conduction and valence bands indices and $M_{k,c,v}$ is the dipolar matrix element. We omitted here the Pauli blocking term for partially filled bands.

\begin{figure}[h!]
    \centering
    \includegraphics[width=7in]{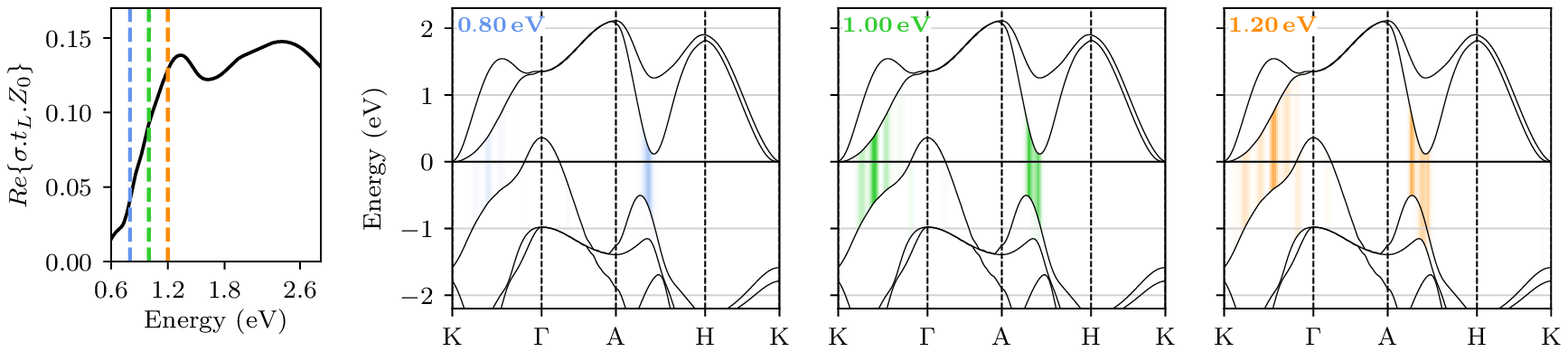}
\vspace{-6ex}
\caption{Contribution diagram of the $\rm{PtSe_2}$ bulk single layer 2D real optical conductivity $\sigma . t_L$ ($\sigma$ is the bulk conductivity, $t_L$ is the layer thickness). The diagram is restricted to the K, $\Gamma$, A and H points where are located the contributing transitions}
    \label{SM-fig:contributionsBulkTail}
\end{figure}

We consider the bulk limit: while the 2D confined model implies a band degeneracy lift upon stacking, we consider here that we operate instead in the continuous $k_z$ limit.
In this picture, the transitions between multiple bands responsible for the low energy absorption tail reduce to transitions between a single pair of bands. The evolution of this coupling is represented along the absorption tail with a contribution diagram in Fig. \ref{SM-fig:contributionsBulkTail}.

We consider a quadratic energy dependency of the electronic transitions involved in the optical absorption, close to the minimum energy: $\omega - \omega_0 \sim (\mathbf{k} - \mathbf{k_0})^2$. We consequently find for the $jDOS$ the standard 3D density of states scaling with frequency:

\begin{equation}
    \begin{split}
        jDOS(\omega)
        &\sim \int \mathrm{d}^3\mathbf{k}\, \delta \left[\omega - E(\mathbf{k})\right] \\
        &\sim \int k^2\mathrm{d}k \,\delta \left[\omega - E(\mathbf{k})\right]\\
        &\sim \int \mathrm{d}E\sqrt{E - \omega_0} \,\delta \left[\omega - E\right]\\
        &\sim \sqrt{\omega - \omega_0}\\
        \label{eq:jDOSCircularMinima}
    \end{split}
\end{equation}

If a single transition is involved then we can consider a constant dipolar matrix element M and use equation (\ref{eq:jDOSCircularMinima}), raising $Re\{\sigma_{2D}\} \sim\omega\,.\,M\,.\,jDOS(\omega) \sim \omega \sqrt{\omega - \omega_0}$. From this dependency one can deduce a linear law to extrapolate from the 2D real optical conductivity the energy $\omega_0$ of the optical transition involved:

\begin{equation}
    \left(\frac{Re\{\sigma_{2D}\}}{\omega}\right)^2 \sim \omega - \omega_0
\end{equation}

\subsection{Rigid shift hypothesis}
\label{SM-sec:rigidShifthyp}

\begin{figure}[h!]
\begin{center}
\includegraphics{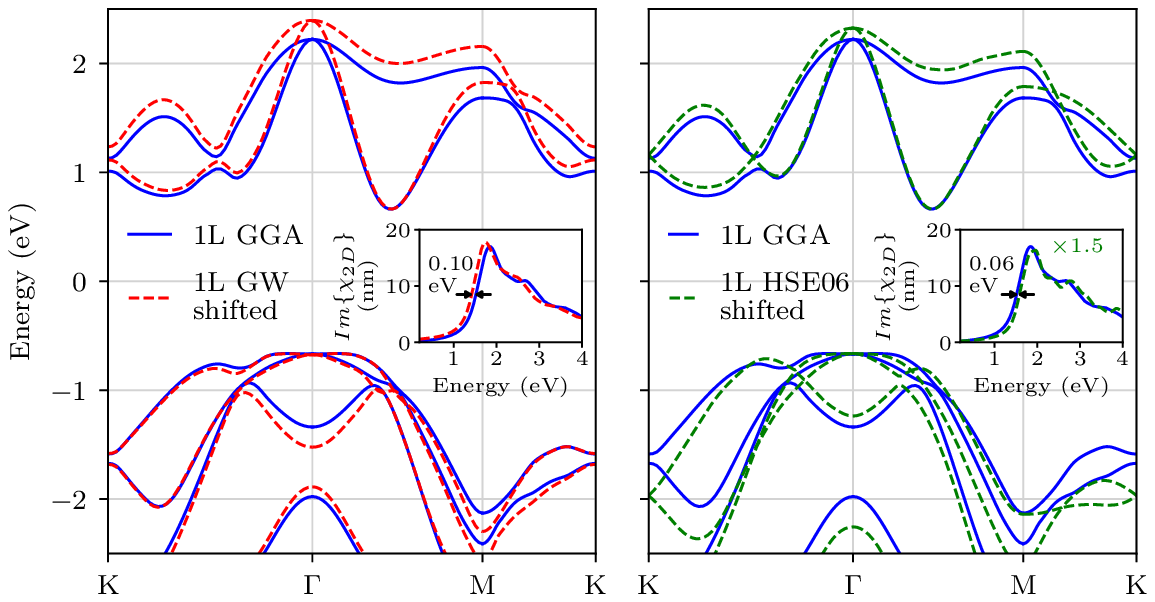}
\end{center}
\vspace{-6.5ex}
\caption{Electronic bands dispersion computed with GGA, GW and HSE06 methods (see section \ref{SM-sec:DFT}). The GW and HSE06 bands are rigidly shifted in order to match the bandgap of the GGA bands dispersion. In inset are represented the imaginary surface susceptibility $Im\{\chi_{2D}\} = Re\{\sigma_{2D}\}\,/\,\varepsilon_0\omega = d\,(Im\{\varepsilon\}-1)$ where the same shifts are applied. A residual a $-0.10\,\mathrm{eV}$ shift of the absorption tail for the GW method and of $+0.06\,\mathrm{eV}$ for the HSE06 tail are present with respect to the GGA absorption tail.}
\label{SM-fig:Bands1LGGAGWHSEShift}
\end{figure}

The hypothesis of a rigid shift of the electronic bands dispersion allows us to correct for the GGA indirect bandgap (Fig. \ref{fig:gapExtrapolation} in the main text). We now justify this hypothesis by comparing the outcome the outcome of different DFT simulations: the GGA method, on which our work is mainly based, as well as GW and HSE methods (sections \ref{sec:GW} and \ref{SM-sec:HSE-RigidShift}). To do so, we apply a rigid shift in between the valence and conduction bands of the GW and HSE06 bands dipersion to set their indirect bandgap value to the one found with the GGA method (Fig. \ref{SM-fig:Bands1LGGAGWHSEShift}).

Comparing the band diagram, we find that the conduction band and the valence band in the vicinity of the $\Gamma$ point appear to be rigidly shifted, but relative distortions can indeed be observed, in particular close to the edge of the first Brillouin zone (M-K). These distrosions have a negligible impact on optical absorption near the direct bandgap, which justifies to neglect them for the study of the near-infrared absorption tail. 

We evaluate the accuracy of the rigid shift hypothesis by evaluating the direct and indirect bandgaps for each method, as well as evaluating their difference (see table \ref{SM-tab:DirectIndirectBandgaps}).
\begin{table}[h!]

    \begin{tabular}{|C{8em}|C{11em}|C{11em}|C{8em}|}
    \hline DFT method  &  Direct bandgap ($\mathrm{eV}$)  & Indirect bandgap ($\mathrm{eV}$) & Difference ($\mathrm{eV}$) \\
    \hline GGA  & 1.53 & 1.33 &  0.20 \\
    \hline GW & 2.79  & 2.56 &  0.23 \\
    \hline HSE06 & 2.00 & 1.83 &  0.17\\
    \hline
    \end{tabular}
        \caption{Direct, indirect bandgaps and differences for GGA, GW and HSE06 DFT methods.}
    \label{SM-tab:DirectIndirectBandgaps}
\end{table}
We find that the energy difference between direct and indirect bandgaps obtained with the various DFT methods are consistent within a $60\,\mathrm{meV}$ spread.

Examining now the optical absorption tail simulated by the different methods (Fig. \ref{SM-fig:Bands1LGGAGWHSEShift}), the shifted GW tail features a remaining $-0.10\,\mathrm{eV}$ shift of the absorption tail with respect to the GGA tail. This shift is of $+0.06\,\mathrm{eV}$ in between the shifted HSE06 optical absorption tail and the GGA-simulated one.  As the GW method features a large bandgap shift ($2\,\mathrm{eV}$) compared to the experimentally extracted one, we consequently expect that this large shift also leads to larger distorsions.

As a consequence, we consider that the characteristic accuracy of the indirect bandgap extraction method used in the main text is of  $60\,\mathrm{meV}$. This a small value against the total bandgap shift ($0.33\,\mathrm{eV}$).
As the indirect bandgap shifts by about $\sim 50\,\rm{meV}\,/\,\mathrm{layer}$ for a thickness of 7 layers, this typical error translates to an estimation range of the semiconductor-to-semimetal transition thickness of 6 to 8 layers.

\section{Photoluminescence of exfoliated flakes} \label{SM-sec:PL}

Photoluminescence is investigated at room temperature, on the visible ($1.2 - 2.5\,\rm{eV}$) range, in order to evaluate the presence of a possible bound exciton in the absorption profile.

\begin{figure}[h!]
\begin{center}
\includegraphics[width=7in]{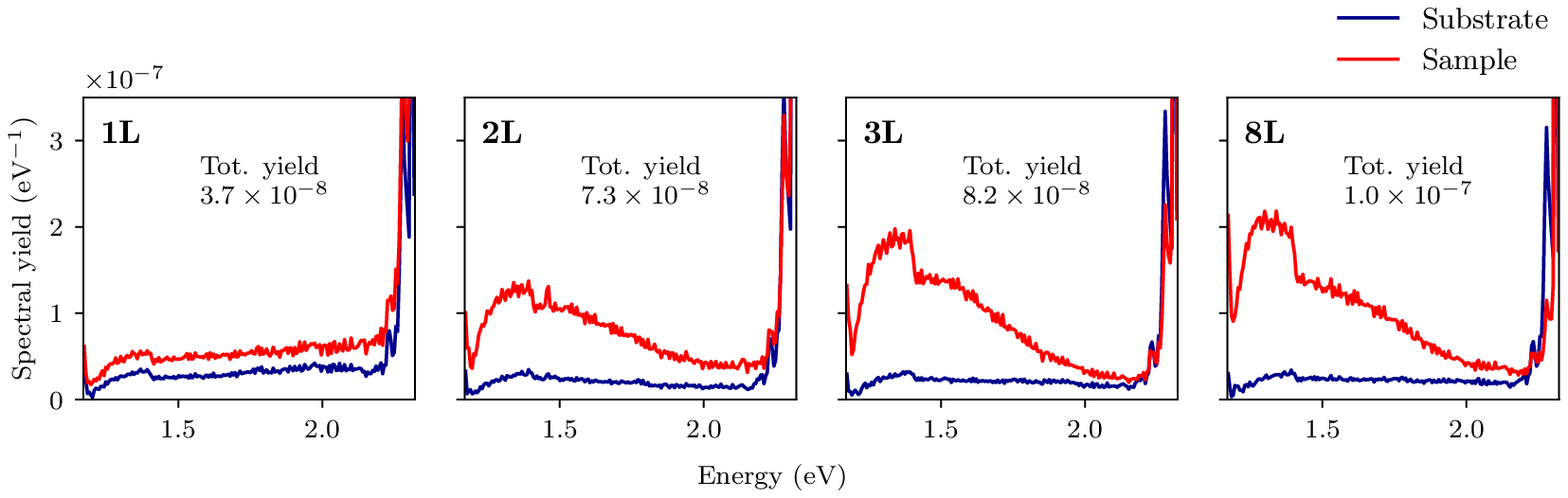}
\end{center}
\vspace{-5ex}
\caption{Photoluminescence signal collected from $\rm{1L}$, $\rm{2L}$, $\rm{3L}$ and $\rm{8L}$ samples and from the substrate at their vicinity, on the $1.2 - 2.3\,\rm{eV}$ range. The signal is normalized and binned as a spectral yield (in unit of $\rm{eV^{-1}}$), and sums up to a total yield, displayed for each sample. The step at $1.4\,\rm{eV}$ is arising from the use of two different spectral windows}
\label{SM-fig:photolum}
\end{figure}

A $0.4\,\rm{mW}$ $532\,\rm{nm}$ green laser light is focused onto selected samples, using a $\times 80$ microscope objective. Photoluminescence from the flake under study and the nearby substrate are collected -- the latter being used as a reference. The measured signal is normalized by the collection efficiency at $570\,\rm{nm}$ and corrected for the CCD and diffraction grating spectral efficiencies. The resulting data is displayed Fig. \ref{SM-fig:photolum}, for 1L, 2L, 4L and 8L samples.

The measured signal is extremely low, with a total yield below $10^{-7}$, pointing to a very inefficient effect. The spectral profile for $\rm{2L}$, $\rm{3L}$ and $\rm{8L}$ sample feature wide bumps around $1.3 - 1.5\,\rm{eV}$, and the one of $\rm{1L}$ appears rather flat (the discontinuity at $1.4\,\rm{eV}$ is an experimental artefact).
These broad photoluminescence peaks can hardly be attributed to excitonic features, but probably originate from some substrate's contamination photoluminescence, exacerbated by the $\rm{PtSe_2}$ absorption at the interface.

\section{Authors contributions}\label{SM-sec:authorsContributions}

MT, RLG, CV and EB designed the experiments with the help of PM. MT conducted the sample fabrication with the help of RLG, JP and MR. MT performed the optical spectroscopy measurements and analysis. SA and MA performed the DFT simulations and atomic orbital projection with the help of RF and FC. SA, MA, and MT computed the optical properties. MT computed the atomic orbital couplings. MT and ED performed the Raman spectroscopy. ED and PL grew the MBE thin film. MT, SA, RLG, BP, FC, CV, RF and EB developed the theoretical interpretation. MT and EB wrote the manuscript with contributions from the coauthors.

\bibliography{ref}

\makeatletter\@input{xx.tex}\makeatother